\documentclass[12pt,epsf]{article}
\usepackage{graphics}
\usepackage{amssymb}
\usepackage{epsf}
\usepackage{epsfig}
\usepackage{rotating}
\usepackage{axodraw}
\usepackage{pstricks}

\def\g{\gamma}

\def\ds#1{#1\kern-1ex\hbox{/}}
\def\dsh{h\kern-1.2ex /}

\newcommand{\bea}{\begin{eqnarray}}
\newcommand{\eea}{\end{eqnarray}}

\def\beq{\begin{equation}}
\def\eeq{\end{equation}}

\def\beqn{\begin{eqnarray}}
\def\eeqn{\end{eqnarray}}
\def\ba{\begin{eqnarray}}
\def\ea{\end{eqnarray}}

\newcommand{\beqa}{\begin{eqnarray}}
\newcommand{\eeqa}{\end{eqnarray}}

\begin{document}
\begin{center}
\vspace{1.cm}
{\bf\large
Unitarity Bounds for Gauged Axionic Interactions and the Green-Schwarz Mechanism \\}
\vspace{1.5cm}
{\bf\large $^{a,b}$Claudio Corian\`{o} $^{a}$Marco Guzzi and $^a$Simone Morelli }

\vspace{1cm}

{\it $^a$Dipartimento di Fisica, Universit\`{a} del Salento \\
and  INFN Sezione di Lecce,  Via Arnesano 73100 Lecce, Italy}\\
\vspace{.5cm}
{\it $^b$ Department of Physics and Institute of Plasma Physics \\
University of Crete, 71003 Heraklion, Greece}\\

\begin{abstract}
We analyze the effective actions of anomalous models in which a four-dimensional version of the
Green-Schwarz mechanism is invoked for the cancellation of the 
anomalies, and we compare it with those models in which gauge
invariance is restored by the presence of a Wess-Zumino term.
Some issues concerning an apparent violation of unitarity of the 
mechanism, which requires Dolgov-Zakharov poles, are carefully 
examined, using a class of amplitudes studied in the past by 
Bouchiat-Iliopoulos-Meyer (BIM), and elaborating
on previous studies.  In the Wess-Zumino
case we determine explicitly the unitarity bound using a realistic model of intersecting branes
(the Madrid model) by studying the corresponding BIM amplitudes.
This is shown to depend significantly on the St\"uckelberg mass 
and on the coupling of the extra anomalous gauge 
bosons and allows one to identify Standard-Model-like 
regions (which are anomaly-free) from regions where the growth of certain amplitudes
is dominated by the anomaly, separated by an inflection 
point which could be studied at the LHC. The bound can even be 
around 5-10 TeV's for a $Z'$ mass around 1 TeV and varies sensitively 
with the anomalous coupling. The results for the WZ case are quite general 
and apply to all the models in which an axion-like interaction is 
introduced as a generalization of  the Peccei-Quinn mechanism, with a gauged axion .
\end{abstract}
\end{center}
\newpage

\section{Introduction}

The cancellation of gauge anomalies in the Standard Model (SM)
is a landmark of modern particle theory that has contributed to
shape our knowledge on the fermion spectrum, its chiral charges 
and couplings. Other mechanisms of cancellation, based on the 
introduction of both local and non-local counterterms, have also 
received a lot of attention in the last two decades, from the 
introduction of the Wess-Zumino term in gauge theories \cite{WZ} 
(which is local) to the Green-Schwarz mechanism of string theory 
\cite{GS} (which is non-local). The field theory realization of 
this second mechanism is rather puzzling also on phenomenological 
grounds since it requires, in four dimensions, the non-local exchange
of a pseudoscalar to restore gauge invariance in the anomalous vertices.
In higher dimensions, for instance in 10 dimensions, the violation of
the Ward identities due to the hexagon diagram is canceled
by the exchange of a 2-form \cite{GS,GSW}.
In this work we are going to analyze the similarities between the
two approaches and emphasize the differences as well. We will try, 
along the way, to point out those unclear aspects of the field theory 
realization of this mechanism - in the absence of supersymmetry and 
gravitational interactions - which, apparently, suffers from the 
presence of an analytic structure in the energy plane that is in 
apparent disagreement with unitarity.
Moving to the WZ case, here we show that the restoration of gauge
invariance in the corresponding
one-loop effective Lagrangian via a local axion counterterm is not
able to guarantee unitarity beyond a certain scale, although this
deficiency is expected \cite{Preskill, CI}, given the local nature
of the counterterm. In the GS case, the restoration of the Ward identities
suffers from the presence of unphysical massless poles in the trilinear gauge
vertices that, as we are going to show, are similar to those present in a
non-local version of axial electrodynamics, which has been studied extensively 
in the past \cite{Adam} with negative conclusions concerning its unitarity
properties.  In particular, in the case of scalar potentials that include
Higgs- axion mixing, the phenomenological interpretation of the GS mechanism
remains problematic in the field theoretical construction.

We comment on the relation between the two mechanisms, when the axion 
is integrated out of the partition function of the anomalous theory, 
and on other issues of the gauge dependence of the
perturbative expansions, which emerge in the different formulations.
In the second part of this work we apply our analysis to a realistic
model characterizing numerically the bounds in effective actions of WZ type and discuss
the possibility to constrain brane and axion-like models at the LHC.

\subsection{WZ and GS counterterms}

Anomalous abelian models are variations of the standard model in which the gauge structure
of this is enlarged by one or more abelian factors. The corresponding anomalies
are canceled by the introduction of a pseudoscalar, an axion ($b$),
that couples to 4-forms $F_I\wedge F_J$  (via $b/M F_I\wedge F_J$, the Wess-Zumino term)
of the gauge fields $(I,J)$ that appear both in ordinary
$(I=J)$ and mixed $(I\neq J)$ anomalies. $M$ is a scale that 
is apparently unrelated to the rest of the theory and
simply  describes the range in which the anomalous model 
can be used as a good approximation to the underlying complete 
theory. The latter can be resolved at an energy $E > M$, 
by using either a renormalizable Lagrangian with an anomaly-free 
chiral fermion spectrum or a string theory. The motivations for 
introducing such models are several, ranging from the study of the 
flavor sector, where several attempts have been performed in the
last decade to reproduce the neutrino mixing matrix using theories
of this type, to effective string models, in which the extra $U(1)$ abound.
We also recall that in effective string models and in models characterized
by extra dimensions the axion ($b$) appears together with a mixing to
the anomalous gauge boson ($\partial{b}B$), which is, by coincidence,
natural in a (Higgs) theory in a broken phase. In a way, theories
of this type have several completions at higher energy \cite{CI}.

Coming to the specific models that we analyze, these are complete
MLSOM-like \cite{CIK,Kiritsis} models with three anomalous $U(1)$
\cite{CIM2,ACG}, while most of the unitarity issues are easier to address
in simple models with two U(1) \cite{CIM1}. In our phenomenological analysis,
which concerns only effective actions of WZ type, we will choose the charge assignments
and the construction of \cite{Ibanez}, but we will work in the region of parameter space
where only the lowest St\"uckelberg mass eigenvalue is taken into account, while the
remaining two extra $Z^{\prime}$ decouple. 
This configuration is not the most general but is enough to
clarify the key physical properties of these models.

\section{Anomaly cancellation and gauge dependences: the GS and the WZ mechanisms in field theory}

In this section we start our discussion of the unitarity properties
of the GS and WZ mechanisms, illustrating
the critical issues. We illustrate a pure diagrammatic 
construction of the WZ effective action using a set of basic 
local counterterms  and show how a certain class of amplitudes 
have an anomalous behavior that grows beyond their unitarity limit at high
energy. The arguments being rather subtle, we have decided to
illustrate the construction of the effective action for both
mechanisms in parallel. A re-arrangement of the same basic
counterterms of the WZ case generates the GS effective action,
which, however, is non-local.  The two mechanisms are different
even if they share a common origin. In a following
section we will integrate out the axion of the WZ formulation to
generate a non-local form of the same mechanism that resembles
more closely the GS counterterm. The two differ by a set of extra
non-local interactions in their respective effective actions.
One could go the other way and formulate the GS mechanism in a
local form using (two or more) extra auxiliary
fields. These points are relevant in order to understand the
connection between the two ways to cancel the anomaly.

\subsection{The Lagrangian}

Specifically, the toy model that we consider 
has a single fermion with a vector-like interaction with the gauge field $A$
and a purely axial-vector interaction with $B$. The Lagrangian is given by
\beqa
\mathcal{L}_0 &=& -\frac{1}{4} F_{A}^{2}
-\frac{1}{4} F_{B}^{2}   + \frac{1}{2}( \partial_{\mu} b +
M_1\ B_{\mu})^{2} + \overline{\psi} i \gamma^{\mu} ( \partial_{\mu} +i e A_{\mu}
+ i g^{}_{B} \gamma^{5} B_{\mu}  ) \psi,
\label{lagrangeBC}
\eeqa
where,  for simplicity, we have taken all the charges to be unitary, and we have allowed for
a St\"uckelberg term for $B$, with $M_1$ being the St\"uckelberg mass
\footnote{Even if (\ref{lagrangeBC}) is not the most general invariant Lagrangian
under the gauge group $U(1)_A\times U(1)_B$, our considerations are the same.
In fact, since $b$ shifts only under a gauge variation of the anomalous $U(1)$
gauge field $B$ (and not under $A$), the gauge invariance of the effective action under a gauge
transformation of the gauge field $A$ requires that there are no terms of the type $b F_A\wedge F_B$.}.
$A$ is massless and takes the role of a photon. The Lagrangian has a
St\"uckelberg-like symmetry with $b\to b -M_1\theta_B$ under a gauge
transformation of $B_\mu$, $\delta B_\mu=\partial_\mu \theta_B$.
The axion is a singlet under gauge transformations of $A$.  We call this simplified
theory the "$A-B$" model. We are allowed not to perform any gauge fixing on $B$
and keep the coupling of the longitudinal component of B to the axion, $\partial B b$,
as an interaction vertex. If we remove $A$, we call the simplified model the "$B$ model".
We will be interchanging between these two models for illustrative purposes and to
underline the essential features of theories of this type.

In the $A-B$ model, the $U(1)_A$ gauge freedom can be gauge-fixed in a generic Lorenz
gauge, with polarization vectors that carry a dependence on the
gauge parameter $\xi_A$, but $A$ being non-anomalous we will
assume trivially the validity of the Ward identities on vector-like
currents. This will erase any dependence on $\xi_A$ both of the
polarization vectors of $A$ and of the propagators of the same gauge boson.
At the same time Chern-Simons (CS) interactions such as $A B\wedge F_B$ 
or $A B \wedge F_A$,  which are present if we define triangle diagrams 
with a symmetric distribution of the partial anomalies of each vertex 
both in the {\bf AVV} (axial-vector/vector/vector) and {\bf AAA} 
cases \cite{CIM1,ACG}, can be absorbed by a re-distribution of the anomaly. 
For instance, if we assume vector Ward identities on the $A$ current 
and move the whole anomaly to the axial-vector currents, then the CS
terms can be omitted. The anomalous corrections in the one-loop
effective action are due to triangle diagrams of the form $BAA$
({\bf AVV}, with conserved vector currents)
and  $BBB$ ({\bf AAA} with a symmetric distribution of the anomalies)
which require two WZ counterterms, given in $S_{WZ}$ below,
for anomaly cancellation. Since the analysis of anomalous gauge
theories containing WZ terms has been the subject of various 
analyses with radically different conclusions regarding the issue
of unitarity of these theories, we refer to the original literature
for more details \cite{Adam,Andrianov1,Andrianov2,Fosco}.
Our goal here is to simply stress the relevance of these previous
analyses in order to understand the difference between the WZ and
GS cancellation mechanism and clarify that Higgs- axion
mixing does not find a suitable description within the standard
formulation of the GS mechanism.

\section{Local and non-local formulations}

The GS mechanism is closely related to the WZ mechanism \cite{WZ}.
The latter, in this case,  consists in restoring gauge invariance
of an anomalous theory by introducing a shifting pseudoscalar, an axion,
that couples to the divergence of an anomalous current. It can be formulated starting
from a massive abelian theory and performing a
{\em field-enlarging transformation} \cite{CIM1}
so as to generate a complete gauge invariant model in which
the usual abelian symmetry is accompanied by a shifting axion.
The original Lagrangian of the massive gauge theory is interpreted
as the gauge-fixed case of the field-enlarged Lagrangian.
The gauge variation of the anomalous effective action is compensated
by the WZ term, so as to have a gauge invariant formulation of the model.
We will show next that a theory built in this way has a unitarity bound
that we will be able to quantify. The appearance of the axion in these
theories seems to be an artifact,
since the presence of a symmetry allows one to set the axion to vanish,
choosing a unitary gauge. In brane models, in the presence of a suitable
scalar potential, the axion ceases to be a gauge artifact and cannot be
gauged away, as shown in \cite{CIK}. This point is rather important,
since it shows that the GS counterterm is unable to describe Higgs- axion
mixing, which takes place when the
Peccei-Quinn \cite{PQ,Srednicki} symmetry of the scalar potential is broken.
The reason is quite obvious: the GS virtual axion is a massless exchange
whose presence is just to guarantee the decoupling of the longitudinal
component of the gauge boson from the anomaly and that does not describe
a physical state. But before coming to a careful analysis of this point,
let us discuss the counterterms of the Lagrangian.

In the $A-B$ model the WZ counterterms are

\beq
\mathcal{L}_{WZ}= \frac{C_{AA}}{2! M_1} b F_A\wedge F_A +
\frac{C_{BB}}{2! M_1} b F_B \wedge F_B,
\label{WZ1}
\eeq
which are fixed by the condition of gauge invariance of the Lagrangian. 
The best way to proceed in the analysis of this theory is to work in the 
$R_\xi$ gauge in order to remove the B-b mixing \cite{CIM1}. Alternatively, 
we are entitled to keep the mixing and perform a perturbative expansion of 
the model using the Proca propagator for the massive gauge boson, and treat 
the $b\partial B$ term as a bilinear
vertex. This second approach can be the source of some confusion,
since one could be misled and identify the perturbative expansion obtained 
by using the WZ theory with that of the GS mechanism, which involves, at a 
field theory level, only a re-definition of the trilinear fermionic vertex
with a pole-like counterterm. In the WZ effective action, treated with the
$b-B$ mixing (and not in the $R_\xi$ gauge), similar counterterms appear.
\begin{figure}[t]
{\centering \resizebox*{12cm}{!}{\rotatebox{0}
{\includegraphics{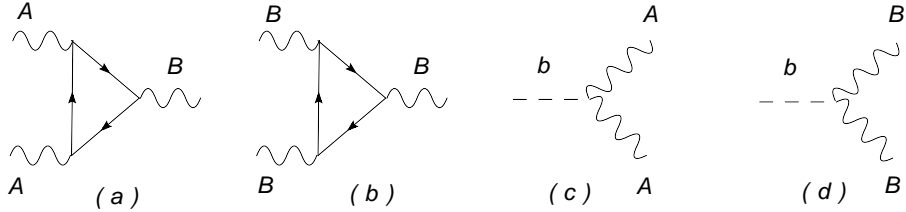}}}\par}
\caption{\small One loop vertices and counterterms in 
the $R_\xi$ gauge for the $A-B$ model for the WZ case.}
\label{onefig}
\end{figure}

\begin{figure}[t]
{\centering \resizebox*{6cm}{!}{\rotatebox{0}
{\includegraphics{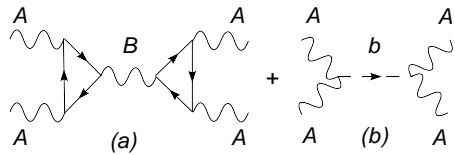}}}\par}
\caption{\small A typical Bouchiat-Iliopoulos-Meyer amplitude and 
the axion counterterm to restore gauge invariance in the $R_\xi$ gauge in the WZ effective action. }
\label{twofig}
\end{figure}
We show in Fig. \ref{onefig} the vertices of the effective action
in the $R_\xi$ gauge approach, and we combine them to describe the process
$A A\to A A$, as shown in Fig.\ref{twofig}.  Graph a) of Fig.\ref{twofig} is a typical
BIM amplitude \cite{BIM}, first studied in '72 by Bouchiat, Iliopoulos and Meyer to
analyze the gauge independence of anomaly-mediated processes in the Standard Model.
The gauge independence of this process is a necessary condition in a gauge
theory in order to have a consistent S-matrix free of spurious singularities
\cite{CIM1}, but is not sufficient to guarantee the absence of a unitarity bound.
Typically, gauge cancellations help to identify the correct power counting
(in $1/M_1$ and in the coupling constants) of the theory and are essential
to establish the overall correctness of the perturbative computations using
the vertices of the effective action.
\begin{figure}[t]
{\centering \resizebox*{12cm}{!}{\rotatebox{0}
{\includegraphics{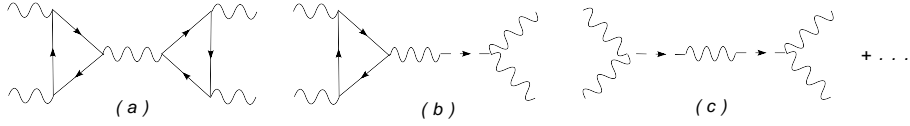}}}\par}
\caption{\small Perturbative expansion of the $A\,A\to A\,A$ amplitude in the presence of $B-b$ mixing. }
\label{fourfig}
\end{figure}
In our example, this can be established as follows:
diagram b) of Fig.\ref{twofig} cancels the gauge dependence of diagram a)
but leaves an overall remnant, which is the contribution of diagram
a) computed in the unitary gauge ($\xi\to \infty$) in which the propagator takes the Proca form
\beq
D_P^{\lambda \lambda'} (k) = - \frac{i}{k^2 - M_1^2} \left( g^{\lambda \lambda'}
- \frac{k^{\lambda} k^{\lambda'}}{M_1^2}\right).
\eeq
Due to this cancellation, the total contribution of the two diagrams is
\beq
\Delta_A^{\lambda\mu\nu}(k,p_1,p_2)D_P^{\lambda \lambda'}
\Delta_A^{\lambda ' \mu' \nu '}(k,k_1,k_2),
\eeq
where $\Delta_A$ is given by the Dolgov-Zakharov parameterization
\ba
\Delta_A(k,p_1,p_2)=A_6(s)(k_{1}+k_{2})^{\lambda}\varepsilon\left[k_1,k_2,\nu,\mu\right]\,,
\ea
where the coefficient $A_6(s)$ in the massless case is $A_6(s)=1/2(\pi^2 s)$.

We call $\Delta_B$ the triangle with a symmetric distribution
of the anomaly ($a_n/3$ for each vertex), which is obtained from $\Delta_A$
by the addition of suitable CS terms \cite{CIM2,ACG}.
The bad behavior of this amplitude at high energy is then trivially given by
\footnote{
We will use the coincise notation $\epsilon[\lambda,p,k,\nu]
\equiv \epsilon^{\lambda\alpha\beta\nu}p_{\alpha}k_{\beta}$ and so on.}

\beq
\frac{1}{M_1^2} \Delta_A^{\lambda \mu\nu}\, \frac{k_\lambda k_{\lambda^{\prime}}}{k^2-M_1^2}
\Delta_A^{\lambda^{\prime}\mu^{\prime}\nu^{\prime}}(k,k_1,k_2) = \frac{1}{M_1^2}
\frac{a_n^2}{k^2 - M_1^2}\epsilon[\mu, \nu, p_1, p_2] \epsilon[\mu^{\prime}, \nu^{\prime}, k_1, k_2],
\eeq
with $a_n= i/(2 \pi^2)$.
Squaring the amplitude, the corresponding cross section  grows linearly with $s=k^2$,
which signals the breaking of unitarity,
as expected in Proca theory, if the corresponding Ward identities are violated.
A similar result holds for the $BBB$ case.
In the alternative  formulation, in which the $b-B$ term is treated as a vertex,
the perturbative expansion is formulated diagrammatically as in Fig. \ref{fourfig}.
Though the expansion is less transparent in this case, it is still
expected to reproduce the results of the $R_\xi$
gauge and of the unitary gauge. Notice that the expansion seems to generate
the specific GS counterterms (Graph 3b)) that
limits the interaction of the gauge field with the anomaly to
its transverse component, together with some extra graphs, which
are clearly not absorbed by a re-definition of the gauge vertex.

\subsection{Integrating out the St\"uckelberg in the WZ case}

\begin{figure}[t]
{\centering \resizebox*{10cm}{!}{\rotatebox{0}
{\includegraphics{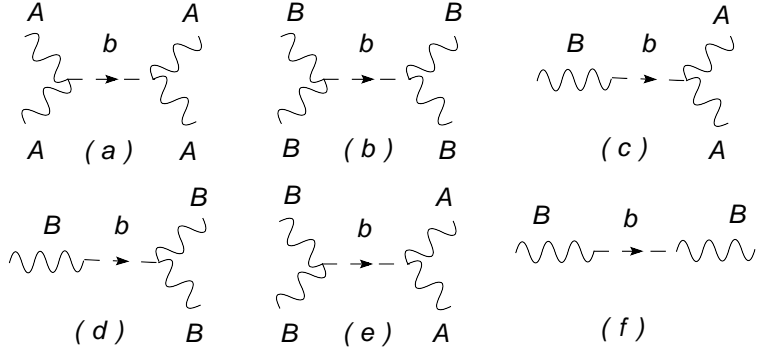}}}\par}
\caption{\small The structure of the WZ effective action having integrated out the axion.}
\label{sixfig}
\end{figure}

\begin{figure}[t]
{\centering \resizebox*{15cm}{!}{\rotatebox{0}
{\includegraphics{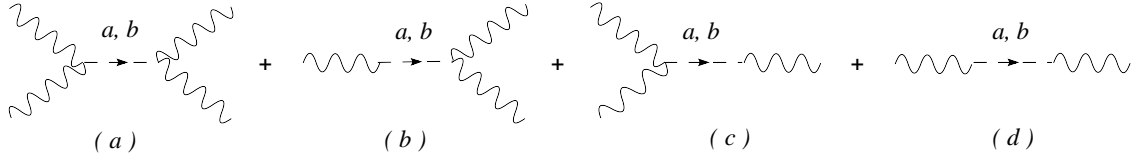}}}\par}
\caption{\small Effective action with two pseudoscalars, one of them ghostlike, $a$. }
\label{feder}
\end{figure}

We can make a forward step and try to integrate out the axion
from the partition function and obtain the non-local version of the WZ
effective action. Notice that this is straightforward only in the case in
which Higgs- axion mixing is absent. The partition function in this case is given by
\beq
Z=\int D\psi D\bar{\psi} D A DB Db \exp \left( i\langle \mathcal{L}(\psi,
\bar{\psi},A,B, b)\rangle\right),
\eeq
where $\langle\, \rangle$ denote integration over $x$ and
\beq
\mathcal{L}=\mathcal{L}_0 + \mathcal{L}_{WZ},
\eeq
with $\mathcal{L}_0$ and $\mathcal{L}_{WZ}$ given in (\ref{lagrangeBC}) and
(\ref{WZ1}), respectively. Indicating with $\mathcal{L}_b$ the $b$ sector of $\mathcal{L}$,
a partial integration on the axion gives
\footnote{We have re-defined
the coefficients in front of the counterterms absorbing the multiplicity factors.}
\beq
\mathcal{L}_b= -\frac{1}{2} b\,\square\, b + b\, J,
\eeq
where
\beq
J= M \partial B - \frac{\kappa_A}{M} F_A\wedge F_A  -
\frac{\kappa_B}{M} F_B\wedge F_B,
\eeq
and performing the path integration over $b$ we obtain
\beq
\int Db \exp\left( i \langle \mathcal{L}\rangle\right)=  \det \left( \square -
M_1^2\right)^{-1/2} \exp\left( \frac{i}{2} J\square^{-1} J\right),
\eeq
where
\beq
\langle J \square^{-1} J \rangle_{WZ} = \langle \left(M_1 \partial B 
-\frac{\kappa_A}{M_1} F_A\wedge F_A  -
\frac{\kappa_B}{M_1} F_B\wedge F_B\right)\square^{-1}
\left(M_1 \partial B - \frac{\kappa_A}{M_1} F_A\wedge F_A -
\frac{\kappa_B}{M_1} F_B\wedge F_B\right)
\rangle.
\eeq
The additional contributions to the effective action are
now non-local and are represented by the set of diagrams in Fig.\ref{sixfig}.
Among these diagrams there are two GS counterterms (diagrams c) and d)),
but there are also other contributions. To generate only the GS counterterms
one needs an additional pseudoscalar called $a$ in order to enforce the cancellation of the extra terms.
There are various ways of doing this \cite{Andrianov1,Andrianov2,Federbush}.
In \cite{Federbush} the non-local counterterm $\partial B \square^{-1} F\wedge F$ of
axial QED, which corresponds to the diagrams b) and c) of Fig.\ref{feder},
is obtained by performing the functional integral over $a$ and $b$ of the following action \cite{Federbush}

\beqa
\mathcal{L} &=& \overline{\psi} \left( i \not{\partial} + e \not{B} \gamma_5\right)\psi - \frac{1}{4} F_B^2 +
\frac{ e^3}{48 \pi^2 M_1} F_B\wedge F_B ( a + b) \nonumber \\
&& + \frac{1}{2}  \left( \partial_\mu b - M_1 B_\mu\right)^2 -
\frac{1}{2} \left( \partial_\mu a - M_1 B_\mu\right)^2.
\label{fedeq}
\eeqa
The integral on $a$ and $b$ are gaussians and their contributions
to the effective action, after integrating them out, are shown in Fig.\ref{feder}.
Notice that $b$ has a positive kinetic term and $a$ is ghost-like. The role of the
two pseudoscalars is to cancel the contributions in Fig. \ref{feder} a) and d), leaving only
the contribution given by graphs b) and c), which has the pole structure typical
of the GS non-local counterterm.
Due to these cancellations, the effective action now reduces to
\beq
 \langle \partial B(x) \square^{-1}(x-y) F(y)\wedge F(y)\rangle
\eeq
besides the anomaly vertex and is represented by interactions of the form b) and c) of Fig. \ref{feder}.
This shows that the WZ and the GS effective actions organize
the perturbative expansions in a rather different way.
It is also quite immediate that the cleanest way to analyze the
expansion is to use the $R_\xi$ gauge, as we have already stressed.
It is then also quite clear that in the WZ case we require the gauge
invariance of the Lagrangian {\em but not} of the trilinear gauge interactions,
while in the GS case, which is realized via Eq. (\ref{fedeq})  or, analogously,
by the Lagrangians proposed in \cite{Andrianov1,Andrianov2}, it is the trilinear
vertex that is rendered gauge invariant (together with the Lagrangian).
The presence of a ghost-like particle in the GS case renders the local
description quite unappealing and for sure the best way to define the mechanism
is just by adding the non-local counterterms. In the WZ case the local description
is quite satisfactory and allows one to treat the $bFF$ interaction
as  a real trilinear vertex, which takes an important role in the presence of
a broken phase.
The GS counterterms are, in practice, the same ones as appearing
in the analysis of axial QED with a non-local counterterm, as we are going to discuss next.
\begin{figure}[t]
{\centering \resizebox*{14cm}{!}{\rotatebox{0}
{\includegraphics{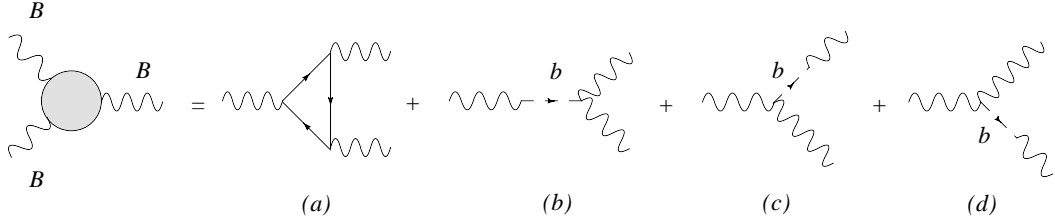}}}\par}
\caption{\small The GS vertex with the non-local 
contributions for the $BBB$ diagram symmetrized on each leg. }
\label{r1}
\end{figure}

\subsection{Non-local counterterms: axial QED}

The use of non-local counterterms to cancel the anomaly is for sure a 
debated issue in quantum field theory since most of the results 
concerning the BRS analysis of these
theories may not apply \cite{Andrianov2}. In the GS case we may ignore all 
the previous constructions and just require ab initio that the
anomalous vertices are modified by the addition of a non-local counterterm
that cancels the anomaly on the axial lines.

Consider, for instance, the case of the $BBB$ vertex of Fig.\ref{r1},
where the regularization of the anomalies has been obtained by adding
the three GS counterterms in a symmetric way
\cite{ABDK}. In the $BAA$ case only a single countertem
is needed, but for the rest the discussion is quite similar to the 
$BBB$ case, with just a few differences. These concern the 
distribution of the partial anomalies on the A and B lines
in the case in which
also $BAA$ is treated symmetrically (equal partial anomalies).
In this particular case we need to compensate the vertex with CS
interactions, which are not, anyhow, observable if the A lines correspond
to conserved gauge currents  such as in QED. In this situation the
Ward identities would force the CS counterterms to vanish. We will
stick to the {\em consistent} definition of the anomaly in which only $B$
carries the total anomaly $a_n$ and $A$ is anomaly-free.
The counterterm used in the GS mechanism both for $BAA$ and $BBB$ is
nothing else but the
opposite of the Dolgov-Zakharov (DZ) expression \cite{Dolgov}, which
in the $BAA$ case takes the form
\ba
C^{\lambda \mu \nu}_{AVV}(k,k_1,k_2) = -\frac{a_n}{k^2} k^\lambda \epsilon[\mu, \nu, k_1, k_2].
\ea
In the BBB case a similar expression is obtained by creating a Bose symmetric combination of DZ poles,
\ba
C^{\lambda \mu \nu}_{AAA}(k,k_1,k_2) = -\left(
\frac{1}{3}\frac{a_n}{k^2} k^\lambda \epsilon[\mu, \nu, k_1, k_2]
+ \frac{1}{3}\frac{a_n}{k_1^2} k_1^\mu \epsilon[\nu, \lambda, k_2, k]
+ \frac{1}{3}\frac{a_n}{k_2^2} k_2^\nu \epsilon[\lambda, \mu, k, k_1]\right).
\ea
We have denoted by $k$ the incoming momenta of the axial-vector vertex and by
$k_1$ and $k_2$ the outgoing momenta of the vector vertices.
We keep this notation also in the AAA case, since $k$ will denote
the momentum exchange in the s-channel when we glue together these 
amplitudes to obtain an amplitude of BIM type; 
this, we will analyze in the next sections.
These expressions are consistent with the following equations of the anomaly for the $BAA$ triangle
\ba
k_{1\mu}C^{\lambda \mu \nu}_{AVV}(k,k_1,k_2)= 0,   \nonumber\\
k_{2\nu}C^{\lambda \mu \nu}_{AVV}(k,k_1,k_2)= 0,   \nonumber\\
k_{\lambda}C^{\lambda \mu \nu}_{AVV}(k,k_1,k_2)= -a_n\epsilon[\mu, \nu, k_1, k_2],   \nonumber\\
\ea
and for the $BBB$ anomalous triangle
\ba
k_{1\mu}C^{\lambda \mu \nu}_{AAA}(k,k_1,k_2)= -\frac{a_n}{3}\epsilon[\lambda, \nu, k, k_2],   \nonumber\\
k_{2\nu}C^{\lambda \mu \nu}_{AAA}(k,k_1,k_2)= -\frac{a_n}{3}\epsilon[\lambda, \mu, k, k_1],   \nonumber\\
k_{\lambda}C^{\lambda \mu \nu}_{AAA}(k,k_1,k_2)=- \frac{a_n}{3}\epsilon[\mu, \nu, k_1, k_2]. \nonumber\\
\ea

So we can define a gauge invariant triangle amplitude, in both the $BBB$ and $BAA$ cases, by
\ba
\Delta^{\lambda \mu \nu\,GS }_{AAA}(k,k_1,k_2) &=& \Delta^{\lambda \mu \nu}_{AAA}(k,k_1,k_2) 
+ C^{\lambda \mu \nu}_{AAA}(k,k_1,k_2) \nonumber\\
\Delta^{\lambda \mu \nu \,GS }_{AVV}(k,k_1,k_2) &=& \Delta^{\lambda \mu \nu}_{AVV}(k,k_1,k_2) 
+ C^{\lambda \mu \nu}_{AVV}(k,k_1,k_2). \nonumber \\
\label{reg1}
\ea
Notice that the (fermionic) triangle diagrams, in the symmetric limit 
$k_1^2=k_2^2=k_3^2$, is exactly the opposite of the DZ counterterms, 
as we will discuss in the next section,
\beq
  \Delta_{AVV}(k_1^2=k_2^2=k^2)=-C_{AVV} \qquad  \Delta_{AAA}(k_1^2=k_2^2=k^2)=-C_{AAA},
\label{reg2}
\eeq
so the cancellation is identical at that point (only at that point), 
and the two vertices $\Delta^{GS}$ vanish. It is rather obvious that the 
cancellations of these poles in BIM amplitudes, corrected by the GS counterterms, 
is identical only for on-shell external gauge lines. It is also
quite straightforward to realize that
the massless $B$-model with the GS vertex correction
is equivalent to axial QED corrected by a non-local term \cite{Adam} 
that is described by the Lagrangian
\beq
\mathcal{L}_{5\,QED}= \overline{\psi} \left( i \not{\partial} 
+ e \not{B} \gamma_5\right)\psi - \frac{1}{4} F_B^2
\eeq
plus the counterterm
\beq
\mathcal{S}_{ct}= \frac{1}{24\pi^2} \langle \partial B(x) \square^{-1}(x-y) F(y)\wedge F(y)
\rangle.
\eeq
This theory is equivalent to the (local) formulation given in 
Eq. \ref{feder} and in \cite{Andrianov2}, where the transversality
constraint ($\partial B=0$) is directly imposed on the Lagrangian via a multiplier.
\begin{figure}[t]
{\centering \resizebox*{4cm}{!}{\rotatebox{0}
{\includegraphics{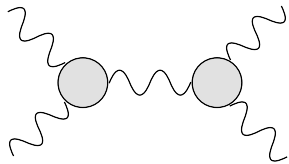}}}\par}
\caption{\small BIM-type amplitude with full GS vertices. 
For on-shell external lines the contributions from the extra poles disappear.   }
\label{eightfig}
\end{figure}

\begin{figure}[h]
{\centering \resizebox*{7cm}{!}{\rotatebox{0}
{\includegraphics{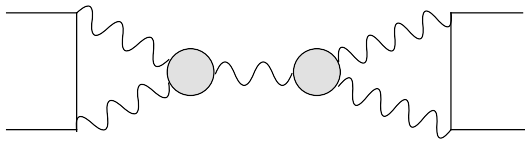}}}\par}
\caption{\small Embedding of the BIM amplitude with GS 
vertices in a fermion/antifermion scattering}
\label{ninefig}
\end{figure}
\begin{figure}[t]
{\centering \resizebox*{12cm}{!}{\rotatebox{0} 
{\includegraphics{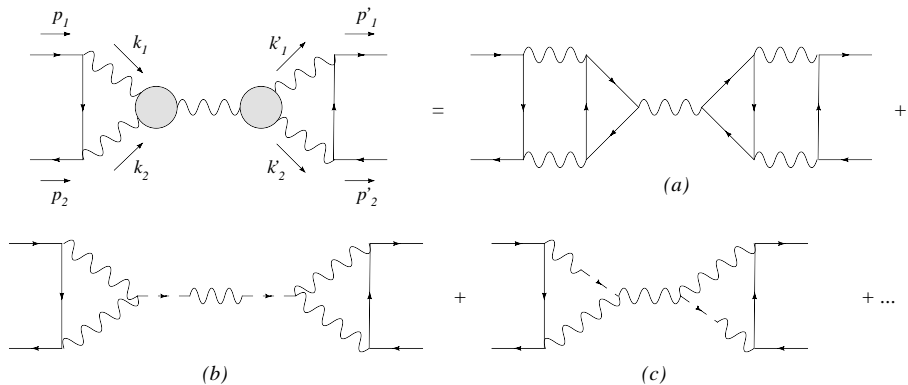}}}\par}
\caption{\small The structure of the fermionic scattering 
amplitudes with spurious massless poles (graphs b) and c)). }
\label{r2}
\end{figure}

Unitarity requires these DZ poles in the C counterterms 
to disappear from a physical amplitude.
To show that this is not the case, in general, consider the 
diagrams depicted in Figs. \ref{eightfig}
and \ref{ninefig}. The structure of the GS vertex
is, for $BBB$, given by (\ref{reg1}) with the three massless poles
generated by the exchange of the pseudoscalar on the three legs, 
as shown in Fig. \ref{r1}.
For on-shell external lines, in this diagram the 
contributions from the extra poles
cancel due to the transversality condition satisfied by the 
polarizators of the gauge bosons.
However, once these amplitudes are embedded into more 
general amplitudes such as those shown
in Fig. (\ref{ninefig},\ref{r2}), the different virtualities 
of the momenta of the anomaly diagrams do not permit,
in general, cancellation of the DZ extra poles introduced
by the counterterm.

We can summarize the issues that we have raised in the following points.

1) The GS and the WZ mechanism have different formulations in terms 
 of  auxiliary fields.
 
2) Previous analyses of axial gauge theories, though distinct
in their Lagrangian formulation, are all equivalent to axial 
QED plus a non-local counterterm. The regularization of the gauge
interactions, in these theories, coincides with that obtained by 
using the GS counterterm on the gauge vertex. In particular, the  
massless poles introduced by the regularization are not understood 
in the context of perturbative unitarity.
 
A special comment is needed when we move to the analysis of Higgs- axion mixing. 
This has been shown
to take place after electroweak symmetry breaking for a special class of potentials,
which are not supersymmetric. The axion, which in the St\"uckelberg phase is essentially
a Goldstone mode, develops a physical component and this component appears as a physical pole.
It is then clear, from the analysis presented above, that the regularization procedure introduced 
by the GS counterterm involves a virtual massless state  and not a physical pole. 
This is at variance with the WZ mechanism, in which  the $bFF$ vertex is introduced 
from the beginning as a vertex and not just as a virtual state. 
In this second case, $b$ can be decomposed in terms of a Goldstone mode and
a physical pseudoscalar, called in \cite{CIK} the {\em axi-Higgs},
which takes the role of a gauged Peccei-Quinn axion \cite{CIK}. This is
entitled to appear as a physical state (and a physical pole, massless or massive) 
in the spectrum. The re-formulation of the GS
counterterm in terms of a pseudoscalar comes also at a cost, due to the presence of
a ghost (phantom) particle in the spectrum, which is absent in the WZ case. There are some
 advantages though, since the theory has, apparently, a nice ultraviolet behavior, 
 given the gauge invariance condition on the vertex.  The presence of these spurious 
 poles requires further investigations to see how they are really embedded into 
 higher order diagrams and we hope to return to this point in the near future in 
 a related work. There is another observation on this issue that is worth to 
 mention: the anomaly is also responsible for a UV/IR conspiracy which is puzzling
 on several grounds. For instance, the linear divergent terms
 $A_1$ and $A_2$ in the Rosenberg representation \cite{Rosenberg} of the anomaly
 diagram are closely related to the infrared anomaly poles in the amplitudes $A_3$
 and $A_6$ in the chiral limit, due to the Ward identity\cite{Alan}.
\footnote{We thank A. R. White for clarifying this point to
us and for describing his forthcoming work on this issue.}

In the next section we will move to the analysis of the unitarity bound
in the WZ  case. As we have shown, in this case it  is possible to
characterize it explicitly. We will work in a specific model, but the
implications of our analysis are general and may be used to constrain 
significantly entire classes of models  containing WZ interactions at
the LHC. Before coming to the specific phenomenological applications we
elaborate on the set of amplitudes that are instrumental in order to
spot the bad high energy behavior of the chiral anomaly
in s-channel processes: the BIM amplitudes.

\section{BIM amplitudes, unitarity and the resonance pole}

The uncontrolled growth of the cross section in the WZ case has
to do with a certain class of amplitudes that have two anomalous
({\bf AVV} or {\bf AAA} ) vertices connected by an s-channel exchange
as in Fig.\ref{fourfig} a). We are interested in the expressions
of these amplitudes in the chiral limit, when all the fermions
are massless. Processes such as $AA\to AA$,  mediated by an
anomalous gauge boson $B$, with on-shell external A lines and
massless fermions, can be expressed in a simplified form, which
is, also in this case, the DZ form. We therefore set
$k_1^2=k_2^2=0$  and $m_f=0$,
which are the correct kinematical conditions to obtain DZ poles.
We briefly elaborate on this point.

We start from the Rosenberg form of the $AVV$ amplitude, which is given by

\ba
&&T^{\lambda\mu\nu}=A_1\varepsilon[k_1,\lambda,\mu,\nu]+A_2\varepsilon[k_2,\lambda,\mu,\nu]
+A_3 k_1^{\mu}\varepsilon[k_1,k_2,\nu,\lambda]
\nonumber\\
&&\hspace{1cm}+A_4 k_2^{\mu}\varepsilon[k_1,k_2,\nu,\lambda]
+A_5 k_1^{\nu}\varepsilon[k_1,k_2,\mu,\lambda]+A_6 k_2^{\nu}\varepsilon[k_1,k_2,\mu,\lambda]\,,
\ea
and imposing the Ward identities we obtain
\ba
&&A_1 = k_2^2 A_4 + k_1\cdot k_2 A_3
\nonumber\\
&&A_2 = k_1^2 A_5 + k_1\cdot k_2 A_6
\nonumber\\
&&A_3(k_1,k_2)=-A_6(k_1,k_2)
\nonumber\\
&&A_4(k_1,k_2)=-A_5(k_1,k_2),
\ea
where the invariant amplitudes $A_3,\dots,A_6$ are free from singularities.
In this specific kinematical limit we can use the following relations to
simplify our amplitude
\ba
&&\varepsilon[k_2,\lambda,\mu,\nu]=\frac{2}{s}
\left(k_2^{\lambda}\varepsilon[k_1,k_2,\nu,\mu]
+k_2^{\nu}\varepsilon[k_1,k_2,\mu,\lambda]
+k_2^{\mu}\varepsilon[k_1,k_2,\nu,\lambda] \right)
\nonumber\\
&&\varepsilon[k_1,\lambda,\mu,\nu]=-\frac{2}{s}
\left(k_1^{\lambda}\varepsilon[k_1,k_2,\nu,\mu]
+k_1^{\nu}\varepsilon[k_1,k_2,\mu,\lambda]
+k_1^{\mu}\varepsilon[k_1,k_2,\nu,\lambda] \right)
\ea
where $k^2=(k_1+k_2)^2=s$ is the center of mass energy.
These combinations allow us to re-write the expression 
of the trilinear amplitude as
\ba
T^{\mu\nu\lambda}=A_6 k^{\lambda}\varepsilon[k_1,k_2,\nu,\mu]+
\left(A_4 + A_6\right)\left(k_2^{\nu}\varepsilon[k_1,k_2,\mu,\lambda]
-k_1^{\mu}\varepsilon[k_1,k_2,\nu,\lambda]\right)\,.
\ea
It is not difficult to see that the second piece drops off 
for physical external on-shell A lines,
and we see that only one invariant amplitude
contributes to the result
\ba
&&T^{\mu\nu\lambda}=A_6^{f}(s)(k_{1}+k_{2})^{\lambda}
\varepsilon\left[k_1,k_2,\nu,\mu\right].
\ea
There are some observations to be made concerning
this result. Notice that $A_6 $ multiplies a longitudinal 
momentum exchange and, as discussed in the literature on 
the chiral anomaly in QCD \cite{Dolgov, Ioffe, Achasov}, 
brings about a {\em massless} pole in $s$. We just recall
that $A_6$ satisfies an unsubtracted dispersion relation 
in $s$ at a fixed invariant mass of the two photons,
$(k_1^2=k_2^2=p^2)$
\beq
A_6(s,p^2)=\frac{1}{\pi^2}\int_{4 m_f^2}^{\infty} dt \, \frac{{\it Im} A_6(t, p^2)}{t - s}
\eeq
and a sum rule 
\beq
\int_{4 m_f^2}^\infty Im A_6(t, p^2) \, d t = \frac{1}{2 \pi},
\eeq
while for on-shell external photons one can use the DZ relation \cite{Dolgov}
\beq
Im A_6(k^2,0)= \frac{1}{\pi} \delta(k^2),
\eeq
to show that the only pole of the amplitude is actually at $s=k^2=0$. 
It can be simplified using the identity
\ba
Li_2(1- a)+Li_2(1- a^{-1})&=&-\frac{1}{2}\log^2(a),
\label{di_log}
\ea
with
\ba
a= \frac{\rho^{}_f + 1}{ \rho^{}_f - 1}  \qquad \rho^{}_f = \sqrt{1 - 4\frac{m_f^2}{s}}\,,
\ea
to give \cite{Achasov}
\ba
&&A_6^{f}(s)=\frac{1}{2\pi^2 s}\left(1 -\frac{m_f^2}{s}\log^{2}\frac{\rho^{}_f + 1 }{\rho^{}_f - 1 }\right).
\ea
We can use this amplitude to discuss both the breaking of unitarity and the cancellation 
of the resonance pole in this simple model. The first point has already been addressed
in the previous section, where the computation of the diagrams in Fig.\ref{twofig}
has shown that only a BIM amplitude survives in the WZ case in the scattering process
$AA\to AA$. If the sum of those two diagrams gives a gauge invariant result, 
with the exchange of the Z' described by a Proca propagator, there is a third 
contribution that should be added to this amplitude. This comes from the exchange 
of the physical axion $\chi$. We recall, in fact, that in the presence of Higgs- axion mixing,
when $b$ is the sum of a Goldstone mode and a physical axion $\chi$, each
anomalous $Z^{\prime}$ is accompanied by the exchange of the $\chi$ \cite{CIK}.
This is generated in the presence of electroweak symmetry breaking, 
having expanded the Higgs scalar $\phi$ around a vacuum $v$
\beq
\phi = \frac{1}{\sqrt{2}}\left( v + \phi^{}_{1} + i \phi^{}_{2} \right),
\eeq
with the axion $b$ expressed as linear combination 
of the rotated fields $\chi^{}_{B}$ and $G^{}_{B}$
\beqa
b = \alpha_1 \chi^{}_{B} + \alpha_2 G^{}_{B} = 
\frac{q^{}_{B} g^{}_{B} v}{M_B} \chi^{}_{B} + \frac{M_1}{M_B} G^{}_{B}.
\eeqa
We also recall that the gauge fields $B^{}_\mu$ get their masses
$M^{}_B$ through the combined Higgs-St\"{u}ckelberg
mechanism,
\beq
M_B=\sqrt{M_1^2 + (q^{}_{B} g^{}_{B} v)^2}.
\eeq
In the phenomenological analysis presented in the next sections the contribution due to $\chi$ has 
been included. Therefore, in the WZ case, the total contributions coming 
from the several BIM amplitudes related to the additional anomalous 
neutral currents should be accompanied not only by the set of Goldstone bosons,
to restore gauge invariance, but also by the exchange of the axi-Higgs.

The cancellation of the resonance pole for $s= M_B$ is an 
important characteristic of BIM amplitudes, which does not occur
in any other (anomaly-free) amplitude. This cancellation is the result 
of some amusingly trivial algebra which we reproduce just for the sake
of clarity. Given a BIM amplitude and Proca exchange, we have
\ba
A_{BIM}&=&   \frac{a_n}{k^2} k^{\lambda} \varepsilon[\mu,\nu,k_1,k_2]  \, \frac{- i}{k^2 - M_1^2}
\left( g^{\lambda \lambda^\prime}
- \frac{k^\lambda k^{\lambda^\prime} }{ M_1^2 }   \right)  \frac{a_n}{k^2}
( - k^{\lambda^\prime}) \varepsilon[\mu^\prime,\nu^\prime,k_1^\prime,k_2^\prime]   \nonumber\\
&=&   \frac{a_n}{k^2}  \varepsilon[\mu,\nu,k_1,k_2]  \, \frac{- i}{ k^2 - M_1^2 }   \frac{k^{ \lambda^\prime}( M_1^2 - k^2 ) }{ M_1^2 }  \,  \frac{a_n}{k^2}
( - k^{\lambda^\prime}) \varepsilon[\mu^\prime,\nu^\prime,k_1^\prime,k_2^\prime]   \nonumber\\
&=&   \frac{a_n}{k^2}  \varepsilon[ \mu,\nu,k_1,k_2]  \, \left(  \frac{- i k^2 }{  M_1^2 }  \right) \,  \frac{a_n}{k^2}
 \varepsilon[\mu^\prime,\nu^\prime,k_1^\prime,k_2^\prime]   \nonumber\\
&=&  - \frac{a_n}{M_1}  \varepsilon[ \mu,\nu,k_1,k_2]  \,\,  \frac{ i }{ k^2 }  \,\,  \frac{a_n}{M_1}
 \varepsilon[\mu^\prime,\nu^\prime,k_1^\prime,k_2^\prime].
\label{resBIM}
\ea
This result implies that the amplitude is described - in the chiral 
limit and for massless external states - by a diagram with the exchange 
of a pseudoscalar (see Fig.\ref{twofig} b)) and that the resonance pole 
has disappeared. It is clear, from (\ref{resBIM}), that these amplitudes 
break unitarity and give a contribution to the cross 
section that grows quadratically
in energy ($\sim s$).  Therefore, searching for
BIM amplitudes at the LHC can be a way to uncover the anomalous
behavior of extra neutral (or charged) gauge interactions.
There is one thing that might tame this growth, and this is
the exchange of the physical axion. We will show, working in
a complete brane model, that the exchange of the $\chi$
does lower the cross section, but insignificantly, independently
from the mass of the axion.

\section{A realistic model with WZ counterterms}

Having clarified the relation between the WZ and GS 
mechanisms using our simple toy-model, we move
towards the analysis of the issue of unitarity 
violation in the WZ Lagrangian at high energy.
Details on the structure of the effective action 
of the complete model that we
are going to analyze can be found in \cite{CIM2}. 
We just mention that this is
characterized by a gauge structure of the form
$SU(3)\times SU(2)\times U(1)\times U(1)_B$, where the $U(1)_B$ is anomalous.
We work in the context of a two-Higgs doublet model with $H_u$ and $H_d$ \cite{CIK}.
In our analysis our setup is that of a complete model, in the sense that
all the charge assignments are those of a realistic brane model with three
extra anomalous $U(1)$, but we will, for simplicity, assume that only
the lowest mass eigenvalue taking part is significant, since the remaining
two additional gauge bosons are heavy and, essentially, decoupled.

We recall that the single anomalous gauge boson, $B$, that we consider in
this analysis is characterized by a generator $Y_B$, which is
anomalous ($TrY_B^3 \neq 0$) but at the same time has mixed anomalies
with the remaining generators of SM and in particular 
with the hypercharge, $Y$. In the presence of a single
anomalous $U(1)$, here denoted as $U(1)_B$, both the $Z$ and the (extra) $Z'$
gauge boson have an anomalous component, proportional to $B$.
We also recall that the effective action of the 
anomalous theory is rendered gauge
invariant using both CS and Green-Schwarz counterterms,
while a given gauge invariant sector involves the exchange both of the
anomalous gauge boson and of the axion  in the $s$-,$t$- and $u$-channels.

In a previous work \cite{ACG} it has been shown how the trilinear vertices of
the effective WZ Lagrangian can be determined consistently for a generic
number of extra anomalous $U(1)$. Here, the goal is to identify and
quantify the contributions that cause a violation of unitarity in this
Lagrangian. For phenomenological reasons it is then convenient to select
those BIM amplitudes that have a better chance, at the experimental level,
to be measured at the LHC and for this reason we will focus on the
process $g\,g \to \gamma \gamma $. The gluon density grows at high
energy especially at smaller (Bjorken variable) $x$-values.  We choose to work with
{\em prompt} final state photons for obvious reasons, the signal being
particularly clean. To begin with, we will be needing the expressions of the
$Z\gamma\gamma$ and the $Z gg$ vertices. In the presence of three anomalous
$U(1)$, here denoted as $U(1)_B$, both the $Z$ and the (extra) $Z'$ gauge boson
have an anomalous component, which is proportional to the $B_{i \mu}$,
the anomalous gauge bosons of the interaction eigenstate basis (i=1,2,3).
The photon vertex is given by \cite{ACG}

\begin{eqnarray}
&&\langle Z_l\g\g\rangle|_{m_f\neq 0}=
-\frac{1}{2}Z^{\lambda}_l A_{\g}^{\mu}A_{\g}^{\nu}
\sum_f\left[
g_Y^3\theta_f^{YYY}\bar{R}^{YYY}_{Z_l\g\g}
+g_2^3\theta_f^{WWW}\bar{R}^{WWW}_{Z_l\g\g}
+g_Y g_2^2\theta_f^{YWW}R^{YWW}_{Z_l\g\g}
\right.\nonumber\\
&&\hspace{2cm}\left.
+ g_Y^2 g_2\theta_f^{YYW}R^{YYW}_{Z_l\g\g}
+\sum_i g_{B_i} g_Y g_2\theta_f^{B_iYW}R^{B_i Y W}_{Z_l\g\g}
\right.\nonumber\\
&&\hspace{2cm}\left.
+\sum_i g_{B_i} g_Y^2\theta_f^{B_iYY}R^{B_iYY}_{Z_l\g\g}
+ g_{B_i} g_2^2\theta_f^{B_iWW}R^{B_iWW}_{Z_l\g\g}\right]
\Delta_{AVV}^{\lambda\mu\nu}(m_f\neq 0).\nonumber\\
\end{eqnarray}
with $l=1,2,3$ enumerating the extra anomalous neutral
currents. The explicit expressions of the rotation matrix $O^A$ can be found in \cite{CIM2,ACG}.
We have defined
\begin{equation}
\bar{R}^{YYY}_{Z_l\g\g}=(O^A)_{YZ_l}(O^{A})_{Y\g}^{2},
\hspace{1cm}
\bar{R}^{WWW}_{Z_l\g\g}=(O^A)_{W_3Z_l}(O^{A})_{W_3\g}^{2},
\end{equation}
and the triangle $\Delta_{AVV}(m_f\neq 0)$ is given by \cite{ACG}
\ba
&&\Delta_{AVV}^{\lambda\mu\nu}(m_f\neq 0,k_1,k_2)=\frac{1}{\pi^2}\int_0^1 dx \int_{0}^{1-x}dy
\frac{1}{\Delta(m_f)}\nonumber\\
&&\hspace{2cm}
\left\{\varepsilon[k_1,\lambda,\mu,\nu]
\left[y(y-1)k_2^2 -x y k_1\cdot k_2\right]
\right.\nonumber\\
&&\hspace{2cm}\left.
+\varepsilon[k_2,\lambda,\mu,\nu]
\left[x(1-x)k_1^2 +x y k_1\cdot k_2\right]
\right.\nonumber\\
&&\hspace{2cm}\left.
+\varepsilon[k_1,k_2,\lambda,\nu]
\left[x(x-1)k_1^{\mu} -x y k_{2}^{\mu} \right]
\right.\nonumber\\
&&\hspace{2cm}\left.
+\varepsilon[k_1,k_2,\lambda,\mu]
\left[x y k_1^{\nu} +(1-y)y k_{2}^{\nu} \right]
\right\}\,,
\nonumber\\
\nonumber\\
&&\Delta(m_f)=m_f^2+x(x-1)k_1^2+y(y-1)k_2^2-2 x y k_1\cdot k_2\,.
\ea
We have defined the following chiral asymmetries
\begin{eqnarray}
&&\theta^{B_{l}YY}_{f}=Q_{B_j,f}^{L}(Q_{Y,f}^{L})^2-Q_{B_l,f}^{R}(Q_{Y,f}^{R})^2,
\nonumber\\
&&\theta^{B_{l}WW}_{f}=Q_{B_l,f}^{L}(T^{3}_{L,f})^2,
\nonumber\\
&&\theta_{f}^{WWW}=(T^{3}_{L,f})^{3},
\nonumber\\
&&\theta_{f}^{YYW}=\left[(Q^{L}_{Y,f})^2 T^{3}_{L,f}\right],
\nonumber\\
&&\theta_{f}^{B_lYW}=\left[Q^{B_l,f}Q^{L}_{Y,f} T^{3}_{L,f}\right],
\end{eqnarray}
with $Q_B^{L/R}$ and $Q_Y^{L/R}$ denoting the charges of the
chiral fermions and $T^3_L$ is the generator of
the third component of the weak isospin, while the $R$ factors
are products of $O^A$ matrix elements.
The matrix $O^A$ relates the interaction eigenstate basis of the 
generators $(Y_B, Y, T_3)$ to those of the
mass eigenstate basis $(T_Z, T_{Z'}, Q)$, of the physical gauge bosons
of the neutral sector, $Z, Z'$ and $A_\gamma$. They are given by
\begin{eqnarray}
&&R^{YYY}_{Z_l\g\g}=3\left[(O^{A})_{Y Z_l}(O^{A})_{Y\g}^{2}\right]\nonumber\\
&&R^{YWW}_{Z_l\g\g}=\left[2(O^{A})_{W_3\g}(O^{A})_{YZ_l}(O^{A})_{Y\g}
+(O^{A})_{W_3\g}^{2}(O^{A})_{Y Z_l}\right]\nonumber\\
&&R^{WWW}_{Z_l\g\g}=\left[3(O^{A})_{B_i Z_l}(O^{A})^{2}_{W_3\g}\right]\nonumber\\
&&R^{YYW}_{Z_l\g\g}=\left[2(O^{A})_{Y Z_l}(O^{A})_{Y\g}(O^{A})_{W_3\g}
+(O^{A})_{W_3 Z_l}(O^{A})_{Y\g}^{2}\right]\nonumber\\
&&R^{B_i YY}_{Z_l\g\g}=(O^{A})_{Y\g}^{2}(O^{A})_{B_i Z_l}\nonumber\\
&&R^{B_i WW}_{Z_l\g\g}=\left[(O^{A})_{W_3\g}^{2}(O^{A})_{B_i Z_l}\right]\nonumber\\
&&R^{B_i YW}_{Z_l\g\g}=\left[2(O^{A})_{B_i Z_l}(O^{A})_{W_3\g}(O^{A})_{Y\g}\right]\,.\nonumber\\
\end{eqnarray}
These expressions will be used extensively in the next section
and computed numerically in a complete brane model.

\subsubsection{MLSOM with one Higgs doublet and an extra singlet}

Another possible framework for the Higgs sector is to consider a Lagrangian
containing one $SU(2)_{W}$ Higgs doublet $H_{u}$
and an extra singlet $\phi$.

In this context all the features concerning the mass matrix for the gauge bosons
remain the same, in fact the covariant derivatives act on the fields as follows
\ba
&&{\cal D}_{\mu}H_{u}=\left(\partial_{\mu} +ig_2\frac{\tau^{i}}{2}W^{i}_{\mu}
+ig_Y q_{u}^{Y}A^{Y}_{\mu}+ig_B \frac{q_{u}^{B}}{2}B_{\mu}\right)H_{u},
\nonumber\\
&&{\cal D}_{\mu}\phi=\left(\partial_{\mu}
+ig_Y q_{\phi}^{Y}A^{Y}_{\mu}+ig_B \frac{q_{\phi}^{B}}{2} B_{\mu}\right)\phi\,,
\ea
where $q_{u}^{B}$ and $q_{\phi}^{B}$ are respectively the charges of the
higgs fields under the extra anomalous $U(1)$ symmetry.

After the spontaneous symmetry breaking, expanding around the vacuum of the two
Higgs bosons, we have
\ba
H^{0}_{u}=v_{u}+\frac{\mbox{Re} H_{u}^{0} +i \mbox{Im}{H_{u}^{0}}}{\sqrt{2}}, &&
\phi=v_{\phi}+\frac{\mbox{Re}{\phi} +i \mbox{Im}{\phi}}{\sqrt{2}}\,.
\ea
By this procedure we obtain a $3\times 3$ mass matrix 
in the mixing of the neutral gauge bosons,
which is similar to that obtained in \cite{CIM2}.

In the Higgs sector the structure of the Peccei-Quinn potential is similar to
that obtained in the presence of two Higgs doublets.
In fact, the symmetric potential is given by
\ba
V_{PQ}=\mu_{u}^2 H_{u}^{\dagger}H_{u} + \lambda_{uu} (H_{u}^{\dagger}H_{u})^2
-2\lambda_{u\phi}(H_{u}^{\dagger}H_{u})\phi^{*}\phi
+ \mu_{\phi}^2 \phi^{*}\phi +\lambda_{\phi\phi}(\phi^{*}\phi)^2\,,
\ea
where the coefficients $\mu_{u}$ and $\mu_{\phi}$ have mass dimension 1 and
$\lambda_{u u},\lambda_{u\phi},\lambda_{\phi\phi}$ are dimensionless,
while the PQ breaking terms can be written as
\ba
V_{\ds{P} \ds{Q}} = \bar{\mu}(H^{\dagger}_{u}H_{u})\phi~e^{i q^{B}_{\phi}\frac{b}{M_1}}
+\bar{\lambda} (\phi^{*}\phi) \phi~e^{i q^{B}_{\phi}\frac{b}{M_1}}
+\textrm{c.c.}\,,
\ea
where $\bar{\mu}$ and $\bar{\lambda}$ have mass dimension 1.
The $CP$-odd sector is still characterized by a $3\times 3$ 
matrix similar to $O^{\chi}$, which we call for
simplicity $O^{\prime\chi}$; this allows for the rotation from 
the interaction eigenstates basis to the physical basis, as follows
\begin{displaymath}
\left(
\begin{array}{c} \mbox{Im} H^{0}_{u} \\
                 \mbox{Im} \phi \\
					  b
\end{array} \right)=O^{\prime\chi}
\left(
\begin{array}{c} \chi\\
                 G^{0}_1\\
					  G^{0}_2
\end{array} \right)\,.
\end{displaymath}

Therefore, in this picture the spectrum in this sector is unchanged, we have a physical axion
and its mass exhibits a combination of two effects, the Higgs vevs
and the presence of a PQ breaking potential.
The parameters can be tuned to be small enough to have a light physical axion.
In general, bounds coming from experimental, astrophysical and cosmological sources give
an upper limit for the axion mass, which is around $10$ meV ~\cite{PDGlive},\cite{Roncadelli}.

\subsection{Prompt photons, the Landau-Yang theorem and the anomaly}

Since the analysis of the unitarity bound will involve the study of 
amplitudes with direct photons in the final state mediated by a $Z$ or
a $Z'$ in the s-channel, we briefly recall some facts concerning the 
structure of these amplitude and in particular the Landau-Yang theorem.
The condition of transversality of the $Z'$ boson ($e_{Z'}\cdot k=0$)
is essential for the vanishing of this
amplitude. A direct proof of the vanishing of the 
on-shell vertex can be found in \cite{CIM1,CIM2}.

The theorem states that a spin 1 particle cannot decay into two
on-shell spin 1 photons because of Bose symmetry and angular momentum conservation.
Angular momentum conservation tells us that the two photons must be
in a spin 1 state (which forces their angular momentum wave function 
to be antisymmetric), while their spatial part is symmetric.
The total wave function is therefore antisymmetric and violates 
the requirement of Bose statistics. For these reasons the amplitude has to vanish.
For a virtual exchange mediated by a $Z'$ the contribution is 
vanishing -after summing over the fermions in each generation of 
the SM-, the theory being anomaly-free, in the chiral limit.
The amplitude is non-vanishing only in the presence of chiral
symmetry breaking terms (fermion masses), which can be induced
both by the QCD vacuum and by the Yukawa couplings of SM in
the presence of electroweak symmetry breaking. For this reason 
it is strongly suppressed also in the SM.

However, the situation in the case of an anomalous vertex
is more subtle. The BIM amplitude is non-vanishing,
but at the same time, as we have explained, is {\em non-resonant}, 
which means that the particle pole due to the $Z'$ has disappeared.
For the rest it will break unitarity at a certain stage.

In fact, a cursory look at the AVV vertex shows that if the external 
photons are on-shell and transverse, the amplitude mediated by this
diagram is proportional to the momentum of the virtual $Z$, $k^{\mu}$. 
This longitudinal momentum
exchange does not set the amplitude to $0$ unless the production
mechanism is also anomalous. We will show first that in the SM these
processes are naturally suppressed, though not identically $0$, since
they are proportional to the fermion masses, due to anomaly cancellation. 
We start our analysis by going back to the AVV diagram, which summarizes
the kinematical behavior of the $Z\gamma \gamma$ amplitude.

Let $k_1$ and $k_2 $ denote the momenta of the two final state photons.
We contract the AVV diagram with the polarization vectors of the photons,
$\varepsilon_{1\mu}$ and $\varepsilon_{2\nu}$ of
the $Z$ boson, $e_{\lambda}$, obtaining
\begin{figure}[t]
{\centering \resizebox*{14cm}{!}{\rotatebox{0}
{\includegraphics{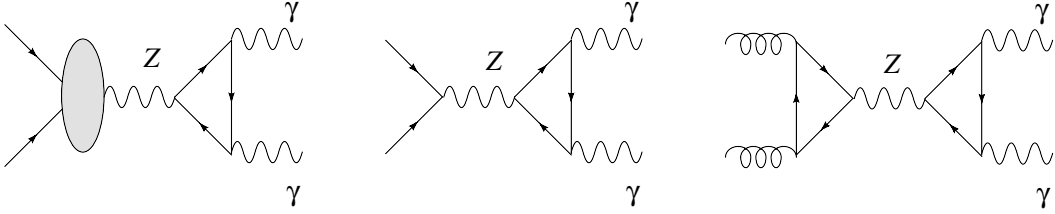}}}\par}
\caption{\small Vanishing and non-vanishing amplitudes mediated by an axial-vector  spin 1. }
\label{VV}
\end{figure}
\ba
&&e_{\lambda}\varepsilon_{1\mu}\varepsilon_{2\nu}
\Delta_{AVV}^{\lambda\mu\nu}(k_1,k_2,m_f\neq 0)=-\frac{1}{\pi^2}\int_0^1 dx \int_{0}^{1-x}dy
\frac{x y}{m_f^2 -2 x y k_1\cdot k_2}\nonumber\\
&&\hspace{2cm}e_{\lambda}\varepsilon_{1\mu}\varepsilon_{2\nu}
\left\{\varepsilon_{\lambda\nu\mu\alpha}
\left( k_{2}^{\alpha}-k_{1}^{\alpha}\right)
+k_1^{\alpha}k_2^{\beta}
\left( \varepsilon_{\alpha\lambda\beta\nu} k_2^{\mu}
-\varepsilon_{\alpha\lambda\beta\mu}k_1^{\nu} \right)
\right\}\,,
\label{Davv}
\ea
where we have used the conditions
\ba
&&k_1^2=k_2^2=0\nonumber\\
&&\varepsilon_{1\mu}k_1^{\mu}=\varepsilon_{2\nu}k_2^{\nu}=0.
\ea
It is important to observe that if we apply Schouten's 
identity we can reduce this expression to the form
\ba
&&e_{\lambda}\varepsilon_{1\mu}\varepsilon_{2\nu}
\Delta_{AVV}^{\lambda\mu\nu}(k_1,k_2,m_f\neq 0)
=e_{\lambda}\left( k_1^{\lambda}+k_2^{\lambda}\right) 
\mathcal{F}_f(k^2,m_f)
\ea
with
\ba
\mathcal{F}_f(k^2, m_f)&=&\mathcal{J}_f(k^2) 
\varepsilon[k_1,k_2,\varepsilon_1,\varepsilon_2] \nonumber \\
\mathcal{J}_f(k^2) &\equiv &-\frac{1}{\pi^2} 
\int_0^1 dx \int_{0}^{1-x}dy \frac{x y}{m_f^2 -2 x y k_1\cdot k_2},
\ea
which vanishes only if we impose the transversality condition on the polarization
vector of the $Z$ boson, $e_{\lambda}k^{\lambda}=0$.  
Alternatively, this amplitude vanishes if the anomalous 
$Z$-photon-photon vertex is contracted with another gauge invariant vertex, as discussed above.
The amplitude has an anomalous behavior. In fact, contracting with the
$k_{\lambda}$ four-vector we obtain
\ba
&&k_{\lambda}\varepsilon_{1\mu}\varepsilon_{2\nu}\Delta_{AVV}^{\lambda\mu\nu}(k_1,k_2,m_f\neq 0)=
\left(-\frac{1}{\pi^2} \int_0^1 dx \int_{0}^{1-x}dy \frac{k_{\lambda}k^{\lambda}x y}
{m_f^2 -2 x y k_1\cdot k_2} \right)
\varepsilon[k_1,k_2,\varepsilon_1,\varepsilon_2]
\nonumber\\
&&\hspace{1cm}=\left(\frac{1}{\pi^2} \int_0^1 dx \int_{0}^{1-x}dy \frac{-2 k_{1}\cdot k_{2}x y}
{m_f^2 -2 x y k_1\cdot k_2} \right)
\varepsilon[k_1,k_2,\varepsilon_1,\varepsilon_2]
\nonumber\\
&&\hspace{1cm}=\left(\frac{1}{\pi^2} + \frac{m_f^2}{\pi^2} \int_0^1 dx \int_{0}^{1-x}dy \frac{1}
{m_f^2 -2 x y k_1\cdot k_2} \right)
\varepsilon[k_1,k_2,\varepsilon_1,\varepsilon_2]
\ea
where the mass-independent and mass-dependent contributions have been separated. Summing over an anomaly-free generation, the first of these two contributions cancel.  

For this reason, it is also convenient to isolate the following quantity
\ba
{\mathcal{G}}_{f}(k^2,m_f)=\frac{m_f^2}{\pi^2} \int_0^1 dx \int_{0}^{1-x}dy \frac{1}
{m_f^2 -2 x y k_1\cdot k_2}
\varepsilon[k_1,k_2,\varepsilon_1,\varepsilon_2]
\ea
which is the only contribution to the triangle amplitude in the SM for a given fermion flavor $f$.

We illustrate in \mbox{Fig.\ref{VV}} the contributions to the 
$q\bar{q}$ annihiliation channel at all orders (first graph), 
at leading order (second graph), and the BIM amplitude (third graph), 
all attached to an $AVV$ final state. In an anomaly-free theory all
these processes are only sensitive to the difference in masses among
the flavors, since degeneracy in the fermion 
mass sets these contributions
to zero. Only the second graph vanishes identically 
due to the Ward identity on the $q\bar{q}$ channel also 
away from the chiral limit. For instance in the third diagram 
chiral symmetry breaking is sufficient to induce violations of 
the Ward identity on the initial state vertex, due to the
different quark masses within a given fermion generation.

To illustrate, in more detail, how chiral symmetry breaking can induce the 
exchange of a scalar component in the process with two prompt photons, we
start from the SM case, where the $Z^* \gamma \gamma$ vertex, multiplied by 
external physical polarizations for the two photons becomes
\ba
V^\lambda_{Z^*\gamma\gamma} = k^\lambda \frac{g_2}{2\cos{\theta}_W} e^2 \sum_f(Q_f)^2
g_{A,f}^{Z}\mathcal{F}_f(k^2, m_f)
\ea
and consider the quark antiquark annihilation channel  
$q\bar{q}\rightarrow Z^{*}\rightarrow\g\g$.
We work in the parton model with massless light quarks. 
We can rewrite the amplitude as
\ba
\mathcal{ M} &=&V^\lambda_{ff Z^*}\Pi_{\lambda,\lambda^{\prime},\xi} V^{\lambda'}_{Z^*\gamma\gamma},
\ea
where we have introduced the $Z q\bar{q}$  vertex
\ba
V_{ff Z}=  \bar{v}(p_1)\Gamma^{\lambda'} u(p_2) \hspace {1cm} \Gamma^\lambda = i g_Z\g^{\lambda}(g_V-g_A\g^{5}),
\ea
and the expression of the propagator of the $ Z $ in the $ R_\xi $ gauge
\ba
\Pi_{\lambda,\lambda^{\prime},\xi}=\frac{-i}{k^2-M_Z^2}\left[g^{\lambda\lambda^{\prime}}
-\frac{k^{\lambda}k^{\lambda^{\prime}}}{k^2-\xi M_Z^2}(1-\xi)\right].
\ea
To move to the unitary gauge, we split the propagator of the $Z $ as
\ba
\Pi_{\lambda,\lambda^{\prime},\xi}=\frac{-i}{k^2-M_Z^2}\left[g^{\lambda\lambda^{\prime}}
-\frac{k^{\lambda}k^{\lambda^{\prime}}}{M_Z^2}\right]
+\frac{-i}{k^2-\xi M_Z^2}\left(\frac{k^{\lambda}k^{\lambda^{\prime}}}{M_Z^2}\right)
\ea
and go to the unitary gauge by choosing $\xi\rightarrow \infty$.  The amplitude will then be written as
\ba
{\cal M} &=& \bar{v}(p_1)\Gamma^\lambda u(p_2)\frac{-i}{k^2-M_Z^2}\left[g^{\lambda\lambda^{\prime}}
-\frac{k^{\lambda}k^{\lambda^{\prime}}}{M_Z^2}\right]
k^{\lambda^{\prime}}\left( \frac{g_2}{2\cos{\theta}_W} e^2 \sum_f(Q_f)^2
g_{A,f}^{Z}\mathcal{F}_f(k^2, m_f) \right)
\nonumber\\
&=&\frac{i}{M_Z^2} \bar{v}(p_1)\Gamma^\lambda u(p_2) k^\lambda  \left( \frac{g_2}{2\cos{\theta}_W} e^2 \sum_f(Q_f)^2
g_{A,f}^{Z}\mathcal{F}_f(k^2, m_f) \right). \nonumber \\
\ea
Clearly, at the Born level, using the Ward identity
on the left $V_{ff Z^*}$ vertex, we find that the
amplitude is zero. This result remains unchanged if we
include higher order corrections (strong/electroweak),
since the structure of this vertex is just modified by a
Pauli (weak-electric) form factor and the additional
contribution vanishes after contraction with the momentum $k^\lambda $.
\begin{figure}[t]
{\centering \resizebox*{14cm}{!}{\rotatebox{0}
{\includegraphics{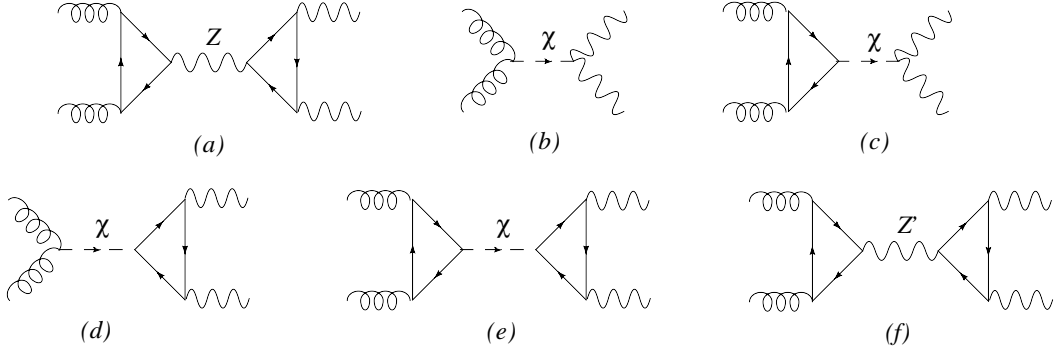}}}\par}
\caption{\small Anomalous contributions in $gg\to \gamma\gamma$  }
\label{GS}
\end{figure}
This amplitude is however non-vanishing if we replace
the $V_{ff Z^*}$ vertex with a $V_{gg Z^*}$ vertex, where now we assume that the
new vertex is computed for non-zero fermion masses (i.e. away from the chiral limit).
In this case we use the Ward identity
\beqn
k_{\rho} \, V_{gg Z^*}^{\rho \nu \mu}  &=&  (p_1 + p_2 )_{\rho} \,  G^{\,  \rho \nu \mu }\nonumber\\
&=& -\frac{ e^2 g_2 }{2 \cos \theta_{W}}  \sum_{q}  g^{Z}_{A,q}  Q^{2}_{q} \,  \epsilon^{\, \nu \mu \alpha \beta }
p_{1 \alpha} p_{2 \beta } \, \left[  \frac{1}{\pi^{2}}  + \frac{m^{2}_{q}}{\pi^{2}} \int^{1}_{0} d x_1 \int^{1 - x_1 }_{0}
d x_2  \,  \frac{1}{ \Delta_q }    \right]
\label{ABJJ_anomaly}
\eeqn
with
\ba
\Delta_q= - x y k^2 +  m_q^2,
\ea

where the constant term $(1/\pi^2)$ vanishes in an anomaly-free theory.
It is convenient to define the function
\ba
\mathcal{G}_q(k^2,m_q)= \epsilon\left[\varepsilon_{1 g},
\varepsilon_{2 g},p_1,p_2\right] m^{2}_{q} \mathcal{I}_q(m_q)
\ea
with
\ba
\mathcal{I}_q(m_q) = \frac{1}{\pi^{2}}\int^{1}_{0} d x_1 \int^{1 - x_1 }_{0} d x_2  \,  \frac{1}{ \Delta_q},
 \ea
where $\varepsilon $ is the polarization of the gluon, which allows one to
express the squared amplitude as
\ba
\langle |\mathcal{M}_{gg\to \gamma\gamma}|^2\rangle &=& \left( \frac{ e^2 g_2 }{2\cos \theta_{W}} \right)^4
 \frac{s^6}{4 M_Z^4} \left( \sum_{q}  g^{q}_{A}  Q^{2}_{q}\mathcal{G}_q(m_q)\right)^2
\left( \sum_{f}  g^{f}_{A}  Q^{2}_{f} \mathcal{G}_f(m_f)\right)^2, \nonumber \\
\ea
with $s\equiv k^2$. Notice that in the large energy limit  
$\mathcal{I}_q\sim \mathcal{J}_f\sim 1/{s^2}$
\cite{Achasov}. This shows that double prompt photon 
production, in the SM, is non-resonant and is proportional
to the quark masses, neglecting the contributions coming 
from the masses of the leptons.

\section{Gauge parameter dependence in the physical basis}

When this analysis is extended to a complete anomalous
model such as the MLSOM
\cite{CIK,CIM2,ACG} even the direct proof of the cancellation
of the gauge dependence
in the $Z'$ exchange is quite complex and not obvious, 
although it is expected at a formal level.
We recall that we are working with a broken phase and the axion
has been decomposed into its physical component ($\chi$, which is the axi-Higgs)
and the Goldstone modes of the extra $Z'$, $G_{Z'}$.
In \cite{ACG} it has been shown that the counterterms
of the theory can be fixed in the St\"uckelberg phase and then re-expressed,
in the Higgs-St\"uckelberg phase, in the physical base.
Although the procedure is formally correct, the explicit check of
cancellation of these gauge dependences is far from trivial and is based on
some identities that we have been able to derive after some efforts, which confirm
the correctness of the approach followed in \cite{CIM2,ACG} for the
determination of the effective Lagrangian of the model after electroweak symmetry breaking.

The matrix $O^{\chi}$, needed to rotate into the mass eigenstates
of the $CP$-odd sector, relating the axion $\chi$ and the
two neutral Goldstones of this sector to the
St\"uckelberg field $b$ and the $CP$-odd phases of the two Higgs doublets
satisfies the following relation

\begin{displaymath}
\left(
\begin{array}{c} ImH^{0}_{u} \\
                 ImH^{0}_{d} \\
					  b
\end{array} \right)=O^{\chi}
\left(
\begin{array}{c} \chi\\
                 G^{0}_1\\
					  G^{0}_2
\end{array} \right)\,,
\end{displaymath}
where the Goldstones in the physical basis are obtained by the following combination
\ba
\label{Goldstones}
G^{Z}&=&G^{0}_1\left[f_u \frac{v_u}{M_Z}O^{\chi}_{12}
+f_d \frac{v_d}{M_Z}O^{\chi}_{22} +g_B \frac{M_1}{M_Z}O^{A}_{ZB}O^{\chi}_{32}\right]
\nonumber\\
&+&G^{0}_2\left[f_u \frac{v_u}{M_Z}O^{\chi}_{13}
+f_d \frac{v_d}{M_Z}O^{\chi}_{23} +g_B \frac{M_1}{M_Z}O^{A}_{ZB}O^{\chi}_{33}\right]
\nonumber\\
&=&c_1 G^{0}_1 +c_2 G^{0}_2\nonumber\\
G^{Z^{\prime}}&=&G^{0}_1\left[f_{u,B} \frac{v_u}{M_Z^{\prime}}O^{\chi}_{12}
+f_{d,B} \frac{v_d}{M_Z^{\prime}}O^{\chi}_{22} +g_B \frac{M_1}{M_Z^{\prime}}
O^{A}_{Z^{\prime}B}O^{\chi}_{32}\right]
\nonumber\\
&+&G^{0}_2\left[f_{u,B} \frac{v_u}{M_Z^{\prime}}O^{\chi}_{13}
+f_{d,B} \frac{v_d}{M_Z^{\prime}}O^{\chi}_{23} +g_B \frac{M_1}{M_Z^{\prime}}
O^{A}_{Z^{\prime}B}O^{\chi}_{33}\right]
\nonumber\\
&=&c^{\prime}_1 G^{0}_1 +c^{\prime}_2 G^{0}_2\,.
\ea
Here we have defined the following coefficients
\ba
f_{u}=g_2 O^{A}_{ZW_3}-g_{Y}O^{A}_{ZY}-q^{B}_{u} g_B O^{A}_{ZB} &&
f_{d}=g_2 O^{A}_{ZW_3}-g_{Y}O^{A}_{ZY}-q^{B}_{d} g_B O^{A}_{ZB}
\nonumber\\
f_{u,B}=g_2 O^{A}_{Z^{\prime}W_3}-g_{Y}O^{A}_{Z^{\prime}Y}-q^{B}_{u} g_B O^{A}_{Z^{\prime}B} &&
f_{d,B}=g_2 O^{A}_{Z^{\prime}W_3}-g_{Y}O^{A}_{Z^{\prime}Y}-q^{B}_{d} g_B O^{A}_{Z^{\prime}B}\,,
\ea
and the $q^{B}_{u,d}$ charges are defined in Table~(\ref{charge_higgs}).
The relations containing the physical Goldstones can be inverted so we obtain
\ba
&&G^{0}_{1}=C_1 G^Z + C_2 G^{Z^{\prime}}\nonumber\\
&&G^{0}_{2}=C_1^{\prime} G^Z +C_2^{\prime} G^{Z^{\prime}}\,,
\ea
where we give the explicit expression only for the coefficient $C^{\prime}_1$,
since this is the one relevant for our purposes.
Then, after the orthonormalization procedure, we obtain
\ba
C^{\prime}_1=\frac{c_2}{\sqrt{c_2^2+c^{\prime 2}_2} }\,.
\ea

\begin{figure}[t]
{\centering \resizebox*{10cm}{!}{\rotatebox{0}
{\includegraphics{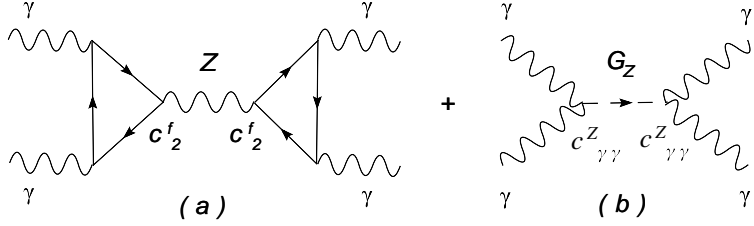}}}\par}
\caption{\small Cancellation of the spurious poles in the physical basis, $m_{f}= 0$.}
\label{gauge_dependence}
\end{figure}

We illustrate the proof of gauge independence from the $Z$ gauge parameter of
the amplitudes in Fig.\ref{gauge_dependence}.
In this case the cancellation of the spurious poles takes place
via the combined exchange of the $Z$ propagator and of the corresponding
Goldstone mode $G_Z$. If we isolate the gauge-dependent
part in the $Z$ boson propagator we obtain

\ba
&&{\cal M}_{A\xi} + {\cal M}_{B\xi}    \nonumber\\
&=& \left( - \frac{1}{2}  c^{f}_{2} \Delta^{\lambda \mu \nu}_{AVV}(p_1, p_2) \right) \frac{- i }{k^2 - \xi_Z M^2_Z } \left( \frac{k^\lambda k^{\lambda^\prime} }{ M^{2}_{Z} } \right)   \left( \frac{1}{2} c^{f}_{2} \Delta^{\lambda^\prime \mu^\prime \nu^\prime}_{AVV}(k_1, k_2) \right)   \nonumber\\
&& + 4 \times \left( 4 c^{Z}_{\g \g} \epsilon[p_1, p_2, \mu, \nu] \right)  \frac{ i }{k^2 - \xi_Z M^2_Z }
 \left( 4 c^{Z}_{\g \g} \epsilon[k_1, k_2, \mu^{\prime}, \nu^{\prime}] \right)     \nonumber\\
&=&  \left( - \frac{a_n}{2}  c^{f}_{2}  \epsilon[p_1, p_2, \mu, \nu]    \right) \frac{- i }{k^2 - \xi_Z M^{2}_{Z} } \left( \frac{1}{ M^2_Z } \right)   \left( \frac{a_n}{2} c^{f}_{2}   \epsilon[k_1, k_2, \mu^{\prime}, \nu^{\prime}]     \right)   \nonumber\\
&& + 4 \times \left( 4 c^{Z}_{\g \g} \epsilon[p_1, p_2, \mu, \nu] \right)  \frac{ i }{k^2 - \xi_Z M^{2}_{Z} }
 \left( 4 c^{Z}_{\g \g} \epsilon[k_1, k_2, \mu^{\prime}, \nu^{\prime}] \right)\,,
\ea
where
\ba
\label{cZ_coeff}
&&c^{Z}_{\g \g}=\left[\frac{F}{M_1}(O^{A}_{W\g})^2+\frac{C_{YY}}{M_1}(O^{A}_{Y\g})^2\right]O^{\chi}_{33}C^{\prime}_1.
\ea
The coefficient $c^{Z}_{\g\g}$ in the ${\cal M}_{B\xi}$  amplitude ($G_Z$ exchange)
must be compared with the massless coefficients
$c^{f}_{2}$ of the ${\cal M}_{A\xi}$ amplitude (Z boson exchange) and
the explicit expressions of the coefficients $C_{YY}$ and $F$ are worked out in the next section.

Adding the contributions of the two diagrams we obtain
\ba
&&{\cal M}_{A\xi} + {\cal M}_{B\xi}    \nonumber\\
&=&  \left\{  \frac{a_n}{2}  \left[ g_B g_Y^2\theta_f^{BYY}R^{BYY}_{Z\g\g}
+ g_B g_2^2\theta_f^{BWW}R^{BWW}_{Z\g\g} \right]  \epsilon[p_1, p_2, \mu, \nu]\right\}
\frac{ i }{k^2 - \xi_Z M^{2}_{Z} } \left( \frac{1}{ M^2_Z } \right)     \nonumber\\
&&\left\{ \frac{a_n}{2} \left[ g_B g_Y^2\theta_f^{BYY}R^{BYY}_{Z\g\g}
+ g_B g_2^2\theta_f^{BWW}R^{BWW}_{Z\g\g}\right]
\epsilon[k_1, k_2, \mu^{\prime}, \nu^{\prime}] \right\}   \nonumber\\
&& + 4 \times \left\{ 4 \left[\frac{F}{M_1}(O^{A}_{W\g})^2
+\frac{C_{YY}}{M_1}(O^{A}_{Y\g})^2 \right] O^{\chi}_{33}C^{\prime}_1
\epsilon[p_1, p_2, \mu, \nu] \right\}
\frac{ i }{k^2 - \xi_Z M^{2}_{Z} }    \nonumber\\
&& \left\{ 4  \left[\frac{F}{M_1}(O^{A}_{W\g})^2+\frac{C_{YY}}{M_1}(O^{A}_{Y\g})^2\right]
O^{\chi}_{33}C^{\prime}_1 \epsilon[k_1, k_2, \mu^{\prime}, \nu^{\prime}] \right\}.
\ea
At this point, the pattern of cancellation can be separated
in three different sectors, a pure $BYY$ sector, a pure BWW and
mixed $BYY$-$BWW$ sectors, and it requires the validity of the relations

\ba
&& \left(  \frac{a_n}{2}  g_B g_Y^2\theta_f^{BYY}R^{BYY}_{Z\g\g}   \right)^2 \frac{1}{M^2_Z} +
4  \left( 4 \frac{C_{YY}}{M_1}(O^{A}_{Y\g})^2  O^{\chi}_{33}C^{\prime}_1 \right)^2     = 0, \\
&& \left(  \frac{a_n}{2}  g_B g_2^2 \theta_f^{BWW}R^{BWW}_{Z\g\g}   \right)^2 \frac{1}{M^2_Z} +
4  \left( 4 \frac{ F }{ M_1 }( O^{A}_{W\g})^2  O^{\chi}_{33}C^{\prime}_1 \right)^2  =  0  ,\\
&& \left( a_n \, g_B g_Y^2 \theta_f^{BYY} R^{BYY}_{Z\g\g}  a_n \, g_B g_2^2 \theta_f^{BWW} R^{BWW}_{Z\g\g}
\right)
\frac{1}{M^2_Z}  +
4 \left( 8 \frac{ F }{ M_1 }( O^{A}_{W\g})^2
\frac{C_{YY}}{M_1}(O^{A}_{Y\g})^2 O^{\chi}_{33}C^{\prime}_1  \right)  = 0.  \nonumber\\
\ea
We have been able to verify that these relations are automatically
satisfied because of the following identity, which connects the rotation
matrix of the interaction to the mass eigenstates $O^A$ to a component
of the matrix $O^{\chi}$. This matrix appears in the rotation from the basis
of St\"uckelberg axions to the basis of the
Goldstones $G_Z$ and $G_{Z'}$ and of the axi-Higgs $\chi$.
The relation is
\ba
O^{A}_{BZ}  \frac{1}{M_Z} = 2 O^{\chi}_{33}C^{\prime}_1 \frac{1}{M_1}\,,
\label{magic_relation_GM}
\ea
with
\ba
O^{\chi}_{33} = \frac{1}{  \sqrt{ \frac{(q^B_u - q^B_d)^2}{M^2_1} \frac{v^2_u v^2_d}{v^2_u+v^2_d }  + 1} },
\ea
with $M_1$ the St\"uckelberg mass.
The origin of this connection has to be found in the 
Yukawa sector and the condition of gauge invariance of the Yukawa couplings.

\section{Unitarity bounds: the partonic contribution $gg\rightarrow \g\g$}

In this section we perform the analytical computation 
of the cross section for the process $gg\rightarrow \g\g$
with the two on-shell gluons ($g$) and the two on-shell photons in the final state.
The same computation is carried out both in the SM and in the MLSOM of \cite{CIK},
where the charge assignments have been determined as in \cite{Ibanez}, to determine
the different behavior of these amplitudes in the two cases. The list of contributions
that we have included are all shown in Fig.\ref{GS}. We report some of the results of the
graphs in order to clarify the notation. For instance we obtain for diagram (a)
\ba
\sigma_A(s)=\frac{1}{2048 \pi} \left[\sum_q \frac{1}{2} c_1^{q}A_{6,q}\right]^2
\left[\sum_{f} \frac{1}{2} c_2^{f}A_{6,f}\right]^2
\frac{s^5}{M_Z^4},
\ea
where we have defined the coefficients
\begin{eqnarray}
&&c_1^{q} = g_3^2 \left[g_Y\theta^{Y}_{q}O^{A}_{YZ}+g_2\theta^{W}_{q}O^{A}_{WZ}
+g_B\theta^{B}_{q}O^{A}_{BZ}\right],
\nonumber\\
&&c_2^{f} = \left[g_Y^3\theta^{YYY}_{f}\bar{R}^{YYY}_{Z\g \g} + g_Y^3\theta^{WWW}_{f}\bar{R}^{WWW}_{Z\g \g}
\right.  \nonumber\\
&& \left. + g_Y g_2^2\theta^{YWW}_{f}R^{YWW}_{Z\g \g}+ g_Y^2 g_2\theta^{YYW}_{f}R^{YYW}_{Z\g \g}
\right.  \nonumber\\
&& \left. +g_B g_Y^2\theta^{BYY}_{f}R^{BYY}_{Z\g \g}+g_B g_2^2\theta^{BWW}_{f}R^{BWW}_{Z\g \g}
\right.  \nonumber\\
&& \left.+g_B g_Y g_2\theta^{BYW}_{f}R^{BYW}_{Z\g \g}\right]\,,
\end{eqnarray}
and the mass of the extra $Z^{\prime}$ is expressed as
\ba
m^{2}_{Z^\prime} &=& \frac{1}{4} \left( 2 M^{2}_{1} + g^{2}v^{2} + N_{BB} +
 \sqrt{ \left( 2 M^{2}_{1} - g^{2}v^{2} + N_{BB}  \right)^2 + 4 g^2 x^2_B }\right)\nonumber\\
&\simeq& M^{2}_{1} + \frac{N_{BB}}{2},   \\
N_{BB} &=& \left(  q^{B\,2}_{u} v^{2}_{u} +  q^{B\,2}_{d} v^{2}_{d}  \right) g^{2}_{B},   \qquad
x_{B} =  \left(  q^{B}_{u} v^{2}_{u} +  q^{B}_{d} v^{2}_{d}  \right) g_{B}.
\ea
We have also defined $v=\sqrt{v_u^2+v_d^2}$, where $v_u$ and $v_d$
are the vevs of the two Higgs bosons,
in the scalar potential \cite{CIK} and $g^2=\sqrt{g_Y^2+g_2^2}$.
We recall that $A_6^{f}(s)$ is approximately given by
\ba
A_6^{f}(s)\approx \frac{1}{2\pi^2 s} -\frac{m_f^2}{2 s^2} 
+ O\left(\frac{m_f^2}{s^2}\log\left(\frac{m_f^2}{s^2}\right)\right),
\ea
at large values of $s$. We have seen that the SM contribution, in the
presence of a massive fermion circulating in the loop is suppressed
by a factor that is $O\left(m_f^2/s^2\right)$.
In the case of an anomalous model this contribution becomes
subleading, the dominant one coming from the anomalous parts,
proportional to the chiral asymmetries $\theta_f$
of the anomalous charge
assignments between left-handed and right-handed fermion modes.
The amplitude in diagram (b) is given by
\ba
\sigma_{B}(s)=\frac{2}{\pi}
\left(g^{\chi}_{gg}\right)^2 \left(g^{\chi}_{\g\g}\right)^2
\frac{ s^3}{(s-M_{\chi}^2)^2}\,.
\ea
where, for convenience, we have defined
\ba
\label{GS_coeffs}
&&g^{\chi}_{gg}=\frac{D}{M_1}O^{\chi}_{31}
\nonumber\\
&&g^{\chi}_{\g\g}=\left[\frac{F}{M_1}(O^{A}_{W\g})^2
+\frac{C_{YY}}{M_1}(O^{A}_{Y\g})^2\right]O^{\chi}_{31}.
\nonumber\\
\ea
We have defined the model-dependent parameters \cite{CIM2}
\ba
&&D=ig_B g_3^2 a_n D_B^{(L)},\hspace{1cm} D_B^{(L)}=-\frac{1}{8}\sum_f Q_{B,f}^{L}
\nonumber\\
&&F=ig_B g_2^2 \frac{a_n}{2} D_B^{(L)},
\nonumber\\
&&C_{YY}=ig_B g_Y^2 \frac{a_n}{2} D_{BYY},\hspace{1cm}
D_{BYY}=-\frac{1}{8}\sum_f \left[Q_{B,f}^{L}(Q_{Y,f}^{L})^2-Q_{B,f}^{R}(Q_{Y,f}^{R})^2\right].
\nonumber\\
\ea
Proceeding in a similar way, graph c) gives
\ba
\sigma_C(s)= \frac{1}{32\pi}  \frac{ \,s^3}{(s-M_{\chi}^2)^2} (g^{\chi}_{gg})^{2}
\left[  \sum_{q} C_0(s,m_q)  c^{\chi, q}_{gg} \right]^2
\ea
where we have used Eq.(\ref{di_log}) and we have set
\ba
C_0(s,m_f) = - \frac{m_f}{4\pi^2 s} \log^2{\frac{\rho^{}_f+1}{\rho^{}_f-1}}.
\ea
Using the Yukawa couplings shown in \cite{CIM2} it is convenient 
to define the following coefficients, which will be used below
\ba
&&c^{\chi,q}_{gg}= g^{2}_{3} c^{\chi,q},   \qquad q=u,d   \nonumber\\
&&c^{\chi,f}_{\gamma\gamma}= e^{2} c^{\chi,f},   \qquad q=u,d,\nu, e.
\label{chi_coupling}
\ea
We have used a condensed notation for the flavors, 
with u = \{u, c, t\}, d = \{d, s, b\}, $\nu$ = \{$\nu_{e}$, $\nu_{\mu}$, $\nu_{\tau}$\}
and e = \{ e, $\mu$, $\tau$\}. The couplings of the physical axion to the fermions are given by
\ba
c^{\chi, u} &=& \Gamma^{u}  \frac{i}{\sqrt 2} O^{\chi}_{11} = \frac{m^{}_{u}}{v_u} i O^{\chi}_{11} ,  \qquad
c^{\chi, d} = - \Gamma^{d} \frac{i}{\sqrt 2} O^{\chi}_{21} = - \frac{m^{}_{d}}{v^{}_{d}} i O^{\chi}_{21},    \nonumber\\
c^{\chi, \nu} &=& \Gamma^{\nu}  \frac{i}{\sqrt 2} O^{\chi}_{11} = \frac{m^{}_{\nu}}{v^{}_u} i O^{\chi}_{11} ,  \qquad
c^{\chi, e} = - \Gamma^{e} \frac{i}{\sqrt 2} O^{\chi}_{21} = - \frac{m^{}_{e}}{v^{}_{d}} i O^{\chi}_{21}.
\ea
We have also relied on the definitions of 
$O^{\chi}$ introduced in a previous work \cite{CIM2}
\ba
O^{\chi}_{11}= - \frac{1}{ \frac{-(q^{B}_{u} 
- q^{B}_{d})}{M_1}v^{}_u \sqrt{    \frac{M^2_1}{ (q^{B}_{u} - q^{B}_{d})^2 }
\frac{v^2}{ v^2_u v^2_d } + 1 } },   \qquad
O^{\chi}_{21}= \frac{1}{ \frac{-(q^{B}_{u} 
- q^{B}_{d})}{M_1}v^{}_u \sqrt{    \frac{M^2_1}{ (q^{B}_{u} - q^{B}_{d})^2 }
\frac{v^2}{ v^2_u v^2_d } + 1 } },
\ea
and the fermion masses have been expressed 
in terms of the Yukawa couplings by the relations
\ba
m_u = \frac{v_u \Gamma_u}{\sqrt 2},  \,\,\,\,\,\, m_d = \frac{v_d \Gamma_d}{\sqrt 2},
\,\,\,\,\,\, m_\nu = \frac{v_u \Gamma_\nu}{\sqrt 2}, \,\,\,\,\,\, m_e = \frac{v_d \Gamma_e}{\sqrt 2}.
\ea
\begin{figure}[t]
{\centering \resizebox*{12cm}{!}{\rotatebox{-90}
{\includegraphics{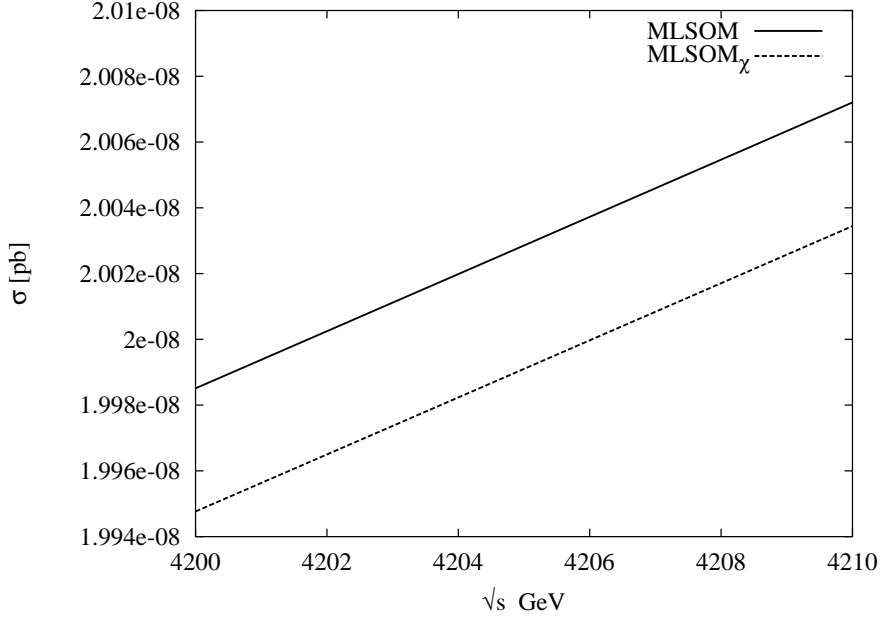}}}\par}
\caption{\small Partonic cross section for the anomalous process $gg\rightarrow \g\g$
with $m_f=0$, $\tan\beta=40$, $g_B=0.1$, $M_{\chi}=10$ GeV and $M_{1}=800$ GeV.
The solid line refers only to the exchange of the $Z$ and the $Z^{\prime}$, while
the dashed line refers to the complete cross section including the $\chi$ exchange.}
\label{MasslessGGgaga}
\end{figure}

\begin{figure}[h]
{\centering \resizebox*{12cm}{!}{\rotatebox{-90}
{\includegraphics{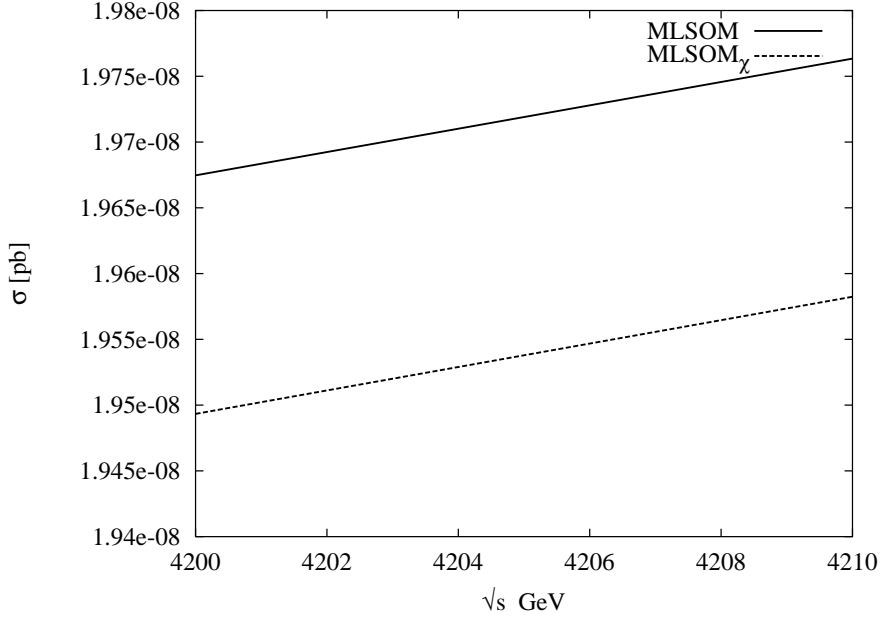}}}\par}
\caption{\small The same as in Fig. \ref{MasslessGGgaga} but with $m_f\neq 0$.}
\label{MassiveGGgaga}
\end{figure}

\begin{figure}[t]
{\centering \resizebox*{12cm}{!}{\rotatebox{-90}
{\includegraphics{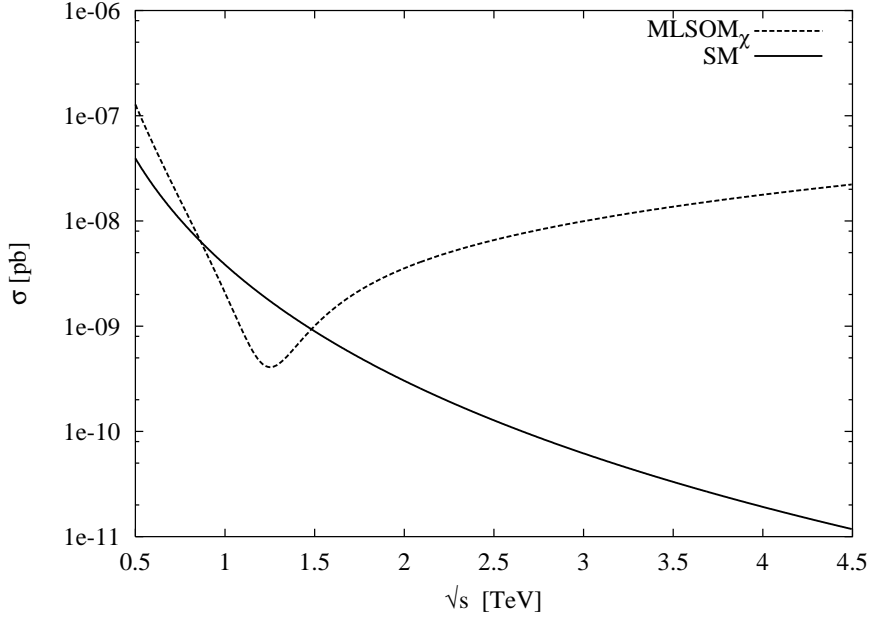}}}\par}
\caption{\small Partonic cross section for the anomalous process 
$gg\rightarrow \g\g$, with $\tan\beta=40$, $g_B=0.1$, $M_{\chi}=10$ GeV and $M_{1}=800$ GeV.
The solid line refers to the SM with the exchange of the $Z$ boson.
The dashed line refers to the MLSOM case. The point of minimum 
divides the anomaly-free region from the region where the anomalous contributions dominate.}
\label{SM_vs_MLSOm}
\end{figure}
The cross section for the amplitude (d) in Fig. (\ref{GS}) is given by
\ba
\sigma_{D}(s) =  \frac{s^{3}}{ 32 \pi (s - M^{2}_{\chi})^{2}} (g^{\chi}_{\g\g})^{2}
 \left[  \sum_{f} C_{0}(s,m_{f})  c^{\chi, f}_{\g\g} \right]^2,
\ea
With these notations, we are now ready to express the cross section for graph e) as
\ba
\sigma_{E}(s) &=& \frac{s^{3}}{2048 \pi (s - M^{2}_{\chi})^{2}}
\left[  \sum_{q} C_{0}(s, m_q) c^{\chi, q}_{gg} \right]^2
\left[  \sum_{f} C_{0}(s, m_{f} ) c^{\chi, f}_{\g\g} \right]^2.
\ea
Finally, the cross section for the $Z^{\prime}$ exchange is given by
\ba
\sigma_F(s)=\frac{1}{2048 \pi} \left[\sum_q \frac{1}{2} {d}_1^{q}A_{6,q}\right]^2
\left[\sum_{f} \frac{1}{2} d_2^{f}A_{6,f}\right]^2
\frac{s^5}{M_{Z^\prime}^4}\,,
\ea
where, again, in order to simplify the notation we have defined the coefficients
\begin{eqnarray}
&&d_1^{q} = g_3^2 \left[g_Y\theta^{Y}_{q} O^{A}_{YZ^\prime}+g_2\theta^{W}_{q} O^{A}_{WZ^\prime}
+g_B\theta^{B}_{q} O^{A}_{BZ^\prime} \right],
\nonumber\\
&&d_2^{f} = \left[g_Y^3\theta^{YYY}_{f}\bar{R}^{YYY}_{Z^\prime \g \g}
+g_2^3\theta^{WWW}_{f}\bar{R}^{WWW}_{Z^\prime \g \g}
\right.  \nonumber\\
&& \left.+ g_Y g_2^2\theta^{YWW}_{f}R^{YWW}_{Z^\prime \g \g}+ g_Y^2 g_2\theta^{YYW}_{f}R^{YYW}_{Z^\prime \g \g}
\right.  \nonumber\\
&& \left. +g_B g_Y^2\theta^{BYY}_{f}R^{BYY}_{Z^\prime \g \g } + g_B g_2^2\theta^{BWW}_{f}R^{BWW}_{Z^\prime \g \g}
\right.  \nonumber\\
&& \left. +g_B g_Yg_2\theta^{BYW}_{f}R^{BYW}_{Z^\prime \g \g }\right].
\end{eqnarray}

\begin{figure}[h]
{\centering \resizebox*{12cm}{!}{\rotatebox{-90}
{\includegraphics{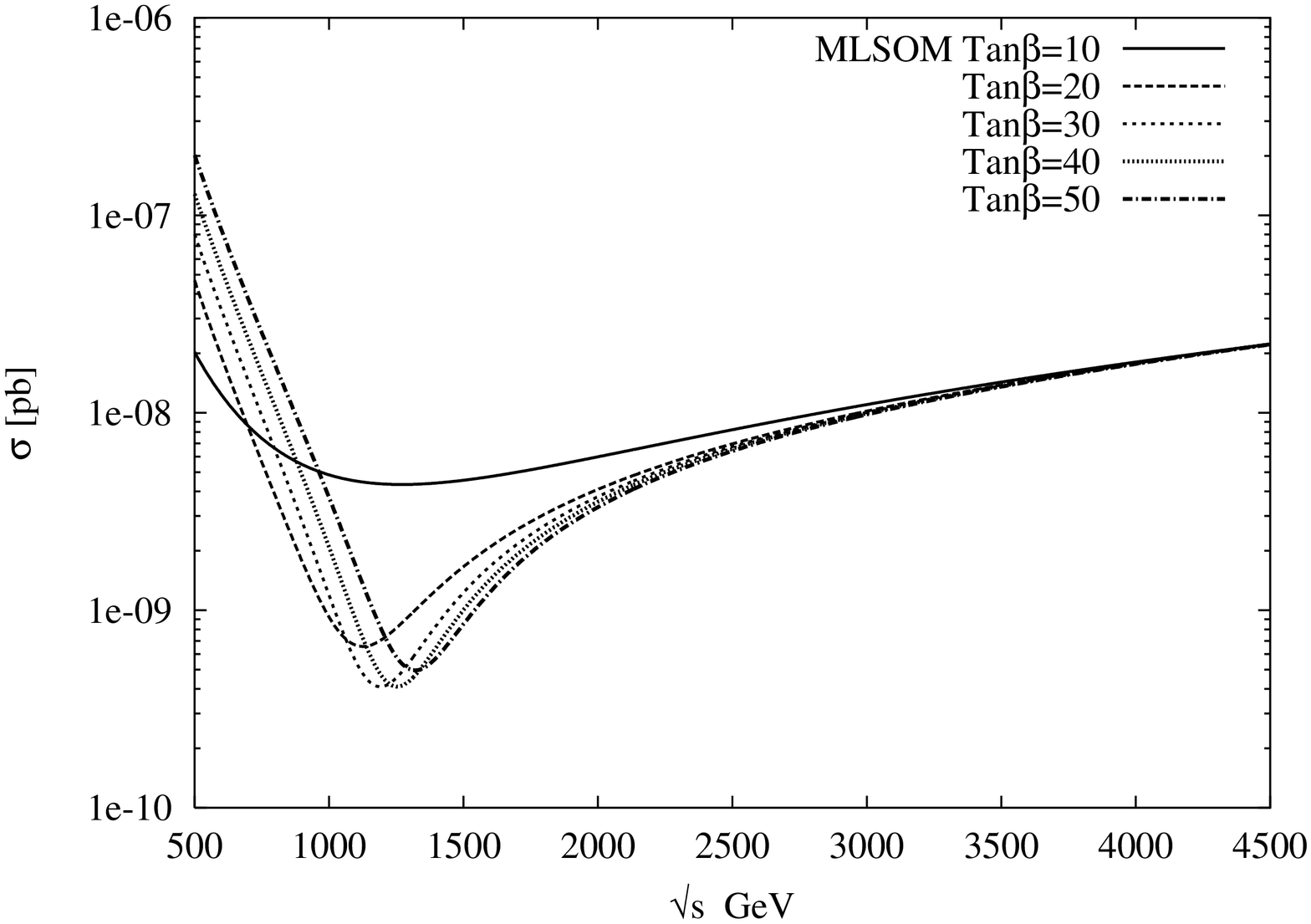}}}\par}
\caption{\small Partonic cross section for the 
anomalous process $gg\rightarrow \g\g$ with $g_B=0.1$, 
$M_{\chi}=10$ GeV and $M_{1}=800$ GeV.
The lines refer to the cross section evaluated 
for different values of $\tan\beta$.}
\label{tanbeta_vs_MLSOm}
\end{figure}

\begin{figure}[t]
{\centering \resizebox*{12cm}{!}{\rotatebox{-90}
{\includegraphics{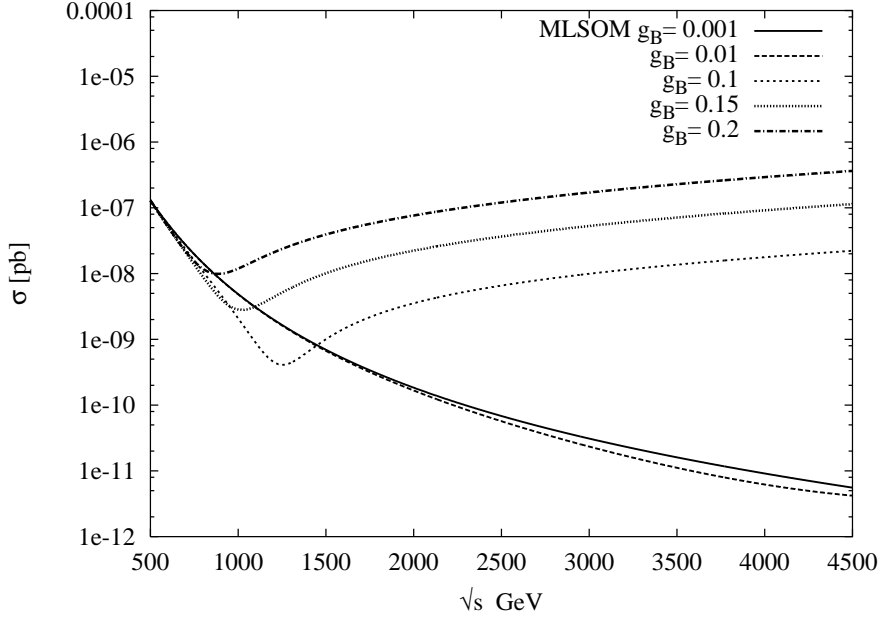}}}\par}
\caption{\small Partonic cross section for the anomalous 
process $gg\rightarrow \g\g$ with $\tan\beta=40$, $M_{\chi}=10$ GeV and $M_{1}=800$ GeV.
The lines refer to the cross section evaluated for different values of the
coupling constant $g_B$.
}
\label{gB_vs_MLSOm}
\end{figure}

\begin{figure}[t]
{\centering \resizebox*{12cm}{!}{\rotatebox{-90}
{\includegraphics{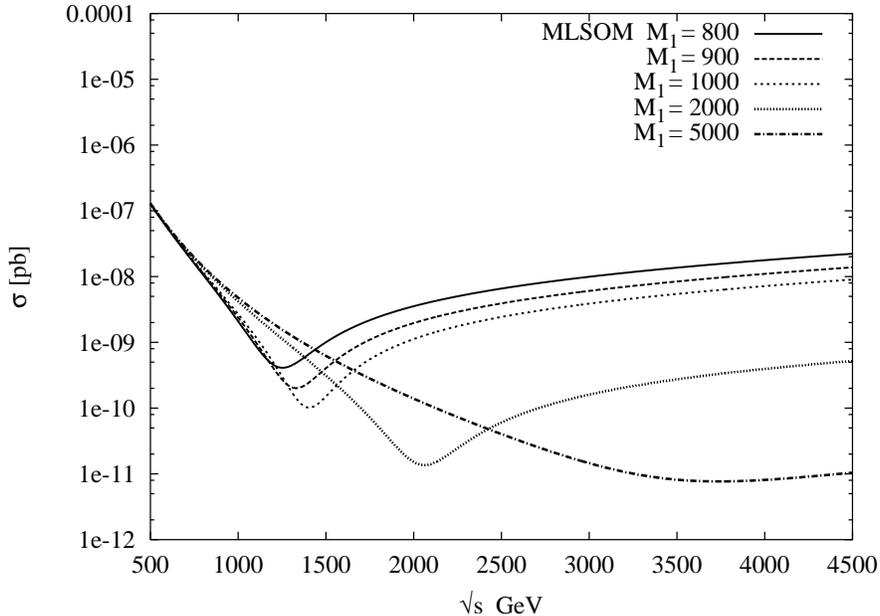}}}\par}
\caption{\small Partonic cross section for the anomalous process $gg\rightarrow \g\g$, with $\tan\beta=40$, $g_B=0.1$ and $M_{\chi}=10$ GeV.
The lines refer to the cross section evaluated for different values of the
St\"uckelberg mass $M_1$.
}
\label{diciotto}
\end{figure}
In the SM case we work in the unitary gauge, 
being a tree-level $Z$ exchange, and we have only
one contribution whose explicit expression is given by
\ba
\sigma^{SM}(s)&=&\frac{1}{32768 \pi^9}\frac{1}{s^3 M_Z^4}
\left[\sum_{q}c_{1,q}^{SM} m^2_q \log^2\left(\frac{\rho_q+1}{\rho_q-1}\right) \right]^{2}
\nonumber\\
&&\times \left[\sum_{f}
c_{2,f}^{SM} m^2_{f} \log^2\left( \frac{\rho_{f} + 1 }{ \rho_{f} - 1 }\right) \right]^{2},
\ea
where we have defined the SM coefficients
\ba
&&c_{1,q}^{SM}= g^{2}_{3}\left[g_Y\theta^{Y}_{q} O^{A}_{YZ}+g_2\theta^{W}_{q}O^{A}_{WZ}\right]
\nonumber\\
&&c_{2,f}^{SM} = \left[ g_Y^3 \theta^{YYY}_{f }\bar{R}^{YYY} + g_2^3\theta^{WWW}_{f }\bar{R}^{WWW}
+ g_Y g_2^2\theta^{YWW}_{f }R^{YWW}+ g_Y^2 g_2\theta^{YYW}_{f }R^{YYW}
\right]\,.\nonumber\\
\ea
The partonic SM cross section is in agreement with unitarity 
in its asymptotic behavior, which is given by

\ba
\sigma^{SM}(s\rightarrow \infty)\approx\frac{\left[ \sum_{q} c_{1,q}^{SM} m^2_q \right]^2 \left[ \sum_{f}
c_{2,f}^{SM} m^2_{f}\right]^2}{s^3 M_Z^4} \, .
\ea

In our anomalous model the complete expression of the same cross section is given by
\ba
&&\sigma^{MLSOM}(s)  \nonumber\\
&&=  \left\{\frac{1}{2048\pi} \frac{s^5}{M_Z^4} 
\left[\frac{1}{2} \sum_q c_1^q \frac{\mathcal{A}(s,m_q)}{2\pi^2 s} \right]^{2}
\left[  \frac{1}{2} \sum_f c_2^f\frac{\mathcal{A}(s,m_f)}{2\pi^2 s}\right]^{2}
 +\frac{2}{ \pi} \frac{\left(g^{\chi}_{gg}\right)^2 \left(g^{\chi}_{\g\g}\right)^2 s^3}
{(s-M_{\chi}^2)^2}\right.  \nonumber \\
&&\left. + \frac{1}{32\pi}  \frac{ (g^{\chi}_{gg})^{2}\,s^3}{(s-M_{\chi}^2)^2}
\left[ \sum_q C_0(s,m_q)  c^{\chi, q}_{gg} \right]^2 +
 \frac{ (g^{\chi}_{\g\g})^{2}   s^{3}}{ 32 \pi (s - M^{2}_{\chi})^{2}}
 \left[ \sum_f C_{0}(s,m_{f})  c^{\chi, f}_{\g\g} \right]^2 
 \right.  \nonumber\\
&& \left.+\frac{s^{3}}{2048 \pi (s - M^{2}_{\chi})^{2}}
\left[ \sum_q  C_{0}(s, m_q) c^{\chi, q}_{gg} \right]^2  
\left[ \sum_f  C_{0}(s, m_{f} ) c^{\chi, f}_{\g\g} \right]^2 \right.  \nonumber\\
&&\left.+ \frac{1}{2048\pi} \frac{s^5}{M_{Z^\prime}^4}
\left[  \frac{1}{2} \sum_q d_1^{\,q}\frac{\mathcal{A}(s,m_q)}{2\pi^2 s} \right]^{2}
\left[  \frac{1}{2} \sum_f  d_2^f\frac{\mathcal{A}(s,m_f)}{2\pi^2 s}\right]^{2}   \right.  \nonumber\\
&&\left. +\frac{ g^{\chi}_{gg} g^{\chi}_{\g\g} s^4}{16\pi M_Z^2(s-M_{\chi}^2)}
\left[ \frac{1}{2} \sum_q c_1^q\frac{\mathcal{A}(s,m_q)}{2\pi^2 s}\right]
\left[ \frac{1}{2} \sum_f c_2^f\frac{\mathcal{A}(s,m_f)}{2\pi^2 s}\right]\right. \nonumber \\
&&+ \left.  \frac{s^{4}}{1024 M^{2}_{Z} \pi (s - M^{2}_{\chi}) }   
\left[ \frac{1}{2} \sum_q  c_{1}^{q} \frac{ \mathcal{A}(s,m_q)}{2\pi^2 s} \right]   
\left[ \frac{1}{2} \sum_f  c_{2}^{f}  \frac{\mathcal{A}(s,m_f)}{2\pi^2 s} \right] \times \right.  \nonumber\\
 && \left. \times \left[ \sum_{f^\prime} C_{0}(s, m_{ f^\prime}) c^{\chi,  f^\prime}_{\g\g}   \right]
\left[  \sum_{q^\prime}   C_{0}(s, m_{ q^\prime}) c^{\chi, q^\prime}_{gg}   \right]   \right.  \nonumber\\
&& \left.+\frac{1}{1024 \pi} \frac{s^5}{M_{Z}^{2}\, M_{Z^\prime}^2}   
\left[ \frac{1}{2} \sum_q {d}_1^{\,q}  \frac{\mathcal{A}(s,m_q)}{2\pi^2 s}   \right]
\left[  \frac{1}{2}  \sum_f d_2^{f}  \frac{\mathcal{A}(s,m_f)}{2\pi^2 s}   \right]  \times   \right.  \nonumber\\
&& \left. \times \left[  \frac{1}{2} \sum_{q^\prime} {c}_1^{q^\prime} 
 \frac{\mathcal{A}(s,m_{q^\prime})}{2\pi^2 s} \right] \left[  \frac{1}{2} \sum_{f^\prime} c_2^{f^\prime}  
 \frac{\mathcal{A}(s,m_{f^\prime})}{2\pi^2 s} \right]  \right.   \nonumber\\
&&  \left. +\frac{s^{4}}{1024\, M^{2}_{Z^\prime}\, \pi \, (s - M^{2}_{\chi}) }   
\left[ \frac{1}{2} \sum_q d_{1}^{\,q}    \frac{\mathcal{A}(s,m_q)}{2\pi^2 s}         \right]   
\left[ \frac{1}{2} \sum_f d_{2}^{f}    \frac{\mathcal{A}(s,m_f)}{2\pi^2 s}  \right]   \times  \right. \nonumber\\
&& \left. \times \left[ \sum_{f^\prime} C_{0}(s, m_{ f^\prime}) c^{\chi,  f^\prime}_{\g\g}   \right]
\left[ \sum_{q^\prime} C_{0}(s, m_{ q^\prime}) c^{\chi, q^\prime}_{gg}   \right]  \right.  \nonumber\\
&&  \left.  +\frac{1}{16\pi}   \frac{  g^{\chi}_{gg}g^{\chi}_{\g\g} \,s^4}{M_{Z^\prime}^2(s-M_{\chi}^2)}
\left[ \frac{1}{2} \sum_q  d_1^{\,q} \frac{\mathcal{A}(s,m_q)}{2\pi^2 s} \right]
\left[ \frac{1}{2} \sum_f d_2^{f} \frac{\mathcal{A}(s,m_f)}{2\pi^2 s} \right]  \right\},
\label{Mchi}
\ea
where we have introduced the notation
\beq
\mathcal{A}(s,m_f)\equiv\left[1-\frac{m^2_f}{s} \log^2\left(\frac{\rho_f+1}{\rho_f-1}\right) \right].
\eeq
At high energy we can neglect the mass of the axion  ($s-M_{\chi}^2\approx s$) 
 and from the limits
\ba
\mathcal{A}(s\rightarrow \infty,m_f)=1,  \qquad C_{0}(s \rightarrow \infty, m_{ f})=0,
\ea
we obtain that the total cross section reduces to
\beq
\sigma^{MLSOM}(s \rightarrow \infty)= \frac{\mathcal{K}^2}{\pi} s,
\eeq
which is linearly divergent and has a unitarity bound.
$\mathcal{K}\equiv \mathcal{K}(s_b,g_B,\alpha_S(s),\tan \beta )$ is defined by
\ba
\mathcal{K}^2=&&   \frac{1}{ 2048 M_Z^4}  \left( \sum_q  \frac{c_{1,q}}{4\pi^2}\right)^2
\left( \sum_f \frac{c_{2,f}}{4\pi^2}\right)^2
+ 2 (g^{\chi}_{gg})^2 (g^{\chi}_{\g\g})^2     \nonumber\\
&&  +\frac{1}{16 M_Z^2} \left( \sum_q \frac{c_{1,q}}{4\pi^2}\right)
\left( \sum_f \frac{c_{2,f}}{4\pi^2}\right)
g^{\chi}_{gg} g^{\chi}_{\g\g} 
+\frac{1}{16 M_{Z^\prime}^2} \left( \sum_q \frac{d_{1,q}}{4\pi^2}\right)
\left( \sum_f \frac{d_{2,f}}{4\pi^2}\right)
g^{\chi}_{gg} g^{\chi}_{\g\g}      \nonumber\\
&& +  \frac{1}{ 1024 M_Z^2 M_{Z^\prime}^2 } \left( \sum_q \frac{c_{1,q}}{4\pi^2}\right)
\left( \sum_f \frac{c_{2,f}}{4\pi^2}\right)
\left( \sum_{q^\prime}  \frac{d_{1,q^\prime}}{4\pi^2}\right) 
\left( \sum_{f^\prime} \frac{d_{2,f^\prime}}{4\pi^2}  \right)    \nonumber\\
&&  + \frac{1}{ 2048 M_{Z^\prime}^4} \left( \sum_q \frac{d_{1,q}}{4\pi^2}\right)^2
\left(  \sum_f \frac{d_{2,f}}{4\pi^2}\right)^2. 
\ea
The derivation of the unitarity bound for this cross section is based, 
in analogy with Fermi theory, on the partial wave expansion
\ba
\frac{d\sigma}{d\Omega}=|f(\theta)|^2=|\frac{1}{2ik}\sum_{l=0}^{\infty}(2l+1)f_l P_l(\cos{\theta})|^2,
\ea
with an $s$-wave contribution given by
\ba
\frac{d\sigma}{d\Omega}=\frac{1}{s}|f_{0}|^2 +\, ...
\ea
Since unitarity requires that  $|f_{l}|\leq 1$ we obtain the bound 

\ba
\frac{d\sigma}{d\Omega}\leq\frac{1}{s}, 
\ea
or, equivalently,
\ba
\sigma\leq\frac{4\pi}{s}
\ea

\ba
\sqrt{s}\geq \sqrt{\frac{2\pi}{\mathcal{K}}}.
\ea
The bound is computed numerically by looking for values $s_b$ at which 
\beq
s_b^2={\frac{2\pi}{\mathcal{K}(s_b,g_B,\alpha_S(s_b),\tan \beta )}}
\eeq
where in the total parametric dependence of the factor ${\mathcal K}$, 
$\mathcal{K}(s_b,g_B,\alpha_S(s_b),\tan \beta )$, we have included the whole energy
dependence, including that coming from running of the coupling (up to three-loop level).
We will analyze below the bound numerically in the context of the specific brane model
of \cite{Ibanez}, \cite{GIIQ}.

\section{Couplings and Parameters in the {\em Madrid model}}

\begin{figure}[t]
{\centering \resizebox*{12cm}{!}{\rotatebox{-90}
{\includegraphics{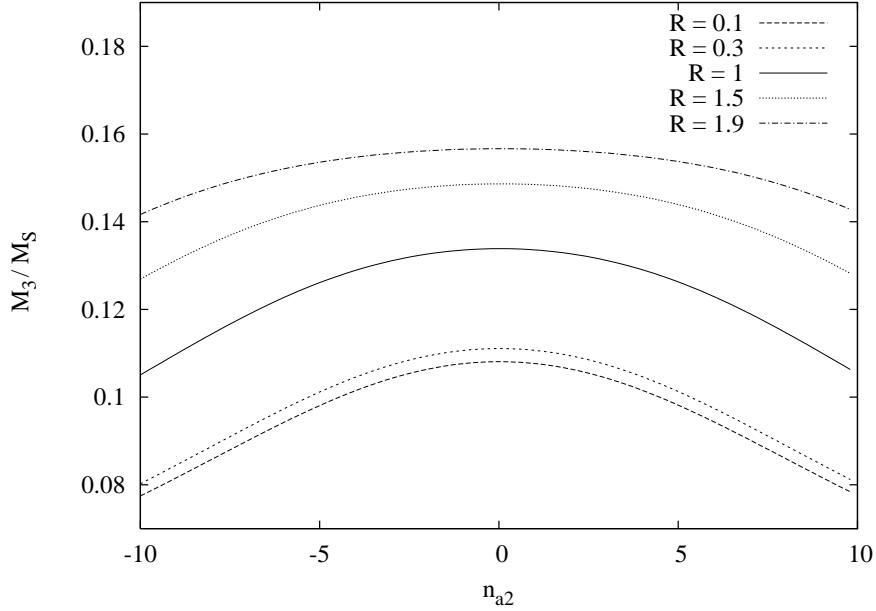}}}\par}
\caption{\small Lightest eigenvalue related to the generator 
$X_B$ for a model of Class A. We have chosen the couplings $g_c$ 
and $g_d$ such that their ratio ${\mathcal{R}}=g_d / g_c$.}
\label{ratio_plot}
\end{figure}

We now turn to a brief illustration of the specific charge assignments of
the class of models that we have implemented in our numerical analysis.
These are defined by a set of free parameters, which can be useful in order
to discern between different scenarios. In our implementation we rotate the
fields from the D-brane basis to the hypercharge basis and at the same time
we redefine the abelian charges and couplings.
The four $U(1)$ in the hypercharge basis are denoted $U(1)_{X_i}$ with
$i=A,B,C$ and $U_Y$, where the last is the hypercharge $U(1)$, which is
demanded to be anomaly-free. This fixes the hypercharge generator
in the hypercharge basis in terms of the generator $q_{\alpha}$ $(\alpha=a,b,c,d)$
in the D-brane basis. The $U(1)_{a}$ and $U(1)_{d}$ symmetries can be identified
with (three times) the baryon number and (minus) the lepton number respectively.
The $U(1)_{c}$ symmetry can be identified with the third component of
right-handed weak isospin and finally the $U(1)_{b}$ is a $PQ$-like symmetry.
Specifically the hypercharge generator is given by
\beq
 Y = \frac{1}{6} (q_a  + 3 q_d )   - \frac{1}{2} q_c,
 \label{constraint}
\eeq
which in fact is a linear combination of the two anomaly-free 
generators $(q_a + 3q_d)$ and $q_c$, while the orthogonal combinations
\beq
X_A =3q_a - q_d ,   \qquad    X_B = q_b,
\eeq
represent anomalous generators in the hypercharge basis. 
Note that relation (\ref{constraint}) must be imposed in 
these models in order to obtain a correct massless hypercharge 
generator as in the SM. The set $(Y,A,B)$ does not depend on 
the model, while the fourth generator $X_C$ is model-dependent and is given by
\beq
X_{C} = \left(  \frac{ 3 \, \beta_2 \, n_{a2} }{ \beta_1 } \, q_a\ +\ 6 \, \rho \, n_{b1} \, q_{b} \ +\
2 \, n_{c1} \, q_c \ +\ \frac{ 3 \, \rho \, \beta_2 \,  n_{d2} }{ \beta_1 } \, q_d  \right).
\eeq
As can be seen in the detailed analysis performed in \cite{Ibanez}
\cite{GIIQ}, the general solutions are parametrized by a phase
$\epsilon =\pm1$, the Neveu-Schwarz background
on the first two tori $\beta_i=1-b_i=1,1/2$, the four integers
$n_{a2},n_{b1},n_{c1}$ and $n_{d2}$, which are the wrapping numbers
of the branes around the extra (toroidal) manifolds 
of the compactification, and a parameter $\rho=1,1/3$,
with an additional constraint in order to obtain the correct massless hypercharge
\beq
n_{c1} \ =\ {{\beta_2}\over {2\beta_1} } (n_{a2} + 3 \, \rho \, n_{d2}).
\label{Y_massless}
\eeq
This choice of the parameters identifies a particular class of models 
which are called Class $A$ models \cite{GIIQ} and all the parameters are
listed in Tables (\ref{parameters},\ref{charge_higgs},\ref{tabpssm}).
Whether anomalous or not, the abelian fields have mass terms induced 
by the St\"uckelberg mechanism, the mass matrix of the $U(1)$ gauge
bosons in the $D$-brane basis is given by the following expression
\beq
({\mathcal{M}}^2)_{\alpha \beta} = g_{\alpha} g_{\beta} M^{2}_{S} 
\sum^{3}_{i=1} c^{\alpha}_{i} c^{\beta}_{i},
\eeq
where $M_{S}$ is some string scale to be tuned. 
Greek indices run over the $D$-brane basis $\{ a,b,c,d\}$, the
Latin index $i$ runs over the three additional abelian gauge groups,
while the $g_{\alpha}$ and $g_{\beta}$ are the couplings of the four $U(1)$.
The eigenvectors $w_i$ (i=Y,A,B,C) and their eigenvalues $\lambda_i$ 
for the matrix $(M^2)_{\alpha \beta}$ have been computed in terms of 
the various classes of models in reference \cite{Ibanez}
\beqa
w^{}_Y &=&   \frac{1}{|w^{}_Y|}  \left\{ \frac{g_d}{3 g_a}, 0, -\frac{g_d}{g_c}, 1  \right\} _{\alpha}    \\
w^{}_{i} &=&  \frac{1}{|w^{}_i|}  \left\{ w_{ia}  , w_{ib} , w_{ic} , 1  \right\}
\eeqa
where the $w_{i=A,B,C}$ are the components of the eigenvectors. 
On the basis of the analysis of the mass matrix one can derive a plot 
of the lightest eigenvalue $\mathcal{M}_3$, which corresponds to the $X_B$
generator in the hypercharge basis \cite{GIIQ}. We have reproduced 
independently the result for the lightest eigenvalue $M_3$, which 
is in good agreement with the predictions of
\cite{GIIQ}. We have implemented numerically the diagonalization 
of the mass matrix and shown in
Fig.\ref{ratio_plot} the behavior of the ratio $M_3/M_S$ as a 
function of the wrapping number $n_{a_2}$ and for several 
values of the ratio $\mathcal{R}=g_d/g_c$. This ratio, which 
characterizes the couplings of $U(1)_c$ and $U(1)_d$, appears 
as a free parameter in the gauge boson mass matrix. $M_S$, the 
string scale, is arbitrary and can be tuned at low values in the region of a few TeV.
 
We have selected a St\"uckelberg mass $M_3$ 
(which is essentially the mass of the extra $Z'$) 
of $800$ GeV. From Fig.\ref{ratio_plot}, if we choose the curve
with $\mathcal{R}=1$ at the peak value, then this values 
is $13\%$ of the string scale, which in this case is 
lowered to approximately 6.1 TeV. It is quite obvious that 
the mass of the extra $Z'$ can be reasonably assumed to be a
free parameter for all practical purposes.

\begin{table}[t]
\begin{center}
\begin{tabular}{|c|c|c|c|c|c|c|c|c|}
\hline
    $\nu$  & $\beta_1$ & $\beta_2 $ & $n_{a2}$  &  $n_{b1}$ & $n_{c1}$ & $n_{d2}$ \\
\hline  1/3 & 1/2  & $  1 $ &  $n_{a2}$ &  -1 & 1 & 1 - $n_{a2}$\\
\hline \end{tabular}
\end{center}
 \caption{\small Parameters for a Class A model with a D6-brane .}
\label{parameters}
\end{table}
 
The matrix $E_{i \alpha}\equiv (w_i)_\alpha$ constructed with the 
eigenvectors of the mass matrix defines the rotation matrix
$U_{i \alpha}$ for the $U(1)$ charges from the D-brane basis $\{ a,b,c,d\}$ into the
hypercharge basis $\{ Y,A,B,C\}$ as follows
\beq
q_{i} = \sum_{\alpha=a,b,c,d} \, U_{i \alpha} q_{\alpha} \,, \,\,\,\,\,  U_{i \alpha} = \frac{g_\alpha}{g_i} E_{i \alpha}, \,\,\,\,\,\,\,\,\,\,\,  (i=Y,A,B,C).
\eeq
So for the hypercharge we find that 
\beq
q_Y = \frac{ g_d }{ 3 g_Y  |w_Y|} (q_a - 3 q_c + 3 q_d)\equiv  \frac{1}{6} (q_a  + 3 q_d )   - \frac{1}{2} q_c,
\eeq 
to be identified with the correct hypercharge assignment given in expression (\ref{constraint}). This identification gives a relation between the gauge couplings
\beq
\frac{1}{g_Y^2} = \frac{|w^{}_Y|^2}{4 g^2_d} = \frac{1}{36 g_a^2} + \frac{1}{4g^2_c} + \frac{1}{4 g^2_d}.
\label{vincolo_gY}
\eeq
For the third generator a similar argument gives
\beq
q^{}_B =   \frac{g_a}{g_B} w^{}_{Ba} \, q_a + \frac{g_b}{g_B} w^{}_{Bb}\, q_b + \frac{g_c}{g_B} w^{}_{Bc}\, q_c + \frac{g_d}{g_B}w^{}_{Bd}\, q_d \equiv q^{}_b,
\eeq
where we identify the gauge symmetry B corresponding to the 
lightest mass eigenvalue as an anomalous generator $q_B$. 
The charges $q_B$ of the SM spectrum are given in Table \ref{charges}.
We recall that given a particular non-abelian $SU(N)$
gauge group, with coupling $g_N$, arising from a stack of $N$ 
parallel branes and the corresponding $U(1)$ field living in the same stack,
the two coupling constants are related by $g_1 = g_N / \sqrt{2 N}$. 
Therefore, in particular the couplings $g_a$ and $g_b$ are determined
using the SM values of the couplings of the non-abelian gauge groups
\beq
g^{2}_{a} = \frac{g^{2}_{QCD}}{6},  \qquad  g^{2}_{b} = \frac{g^{2}_{L}}{4},
\eeq
where $g_L$ is the $SU(2)_L$ gauge coupling, while $g_c$ and $g_d$ 
are constrained by relation (\ref{vincolo_gY}).
Imposing gauge invariance for the Yukawa couplings \cite{CIM2}
we obtain the assignments for the Higgs doublets 
shown in Table (\ref{charge_higgs}).
\begin{table}[t]
\begin{center}
\begin{tabular}{|c|c|c|c|c|}
\hline
   &  Y &$ X^{}_A$  & $X^{}_{B} $    \\
\hline $H^{}_{u}$ & 1/2 & 0  & 2     \\
\hline  $H^{}_{d}$   & 1/2  &0  & -2     \\
\hline \end{tabular}
\end{center}
 \caption{\small Higgs charges in the Madrid model.  }
\label{charge_higgs}
\end{table}
\begin{table}[t]
\begin{center}
\begin{tabular}{|c|c|c|c|c|c|}
\hline
   &   $q_a$  & $q_b $ & $q_c $ & $q_d$   \\
\hline $Q_L$ & 1  & -1 & 0 & 0   \\
\hline  $u_R$   &  -1  & 0  & 1  & 0   \\
\hline   $d_R$  &  -1  & 0  & -1  & 0   \\
\hline  $ L$    &  0   & -1   & 0  & -1    \\
\hline   $e_R$  &  0  & 0  & -1  & 1      \\
\hline   $N_R$  &  0  & 0  & 1  & 1    \\
\hline \end{tabular}
\end{center}
 \caption{\small SM spectrum charges in the $D$-brane basis for the Madrid model.}
\label{tabpssm}
\end{table}
\begin{table}[t]
\begin{center}
\begin{tabular}{|c|c|c|c|c|c|c|}
\hline
   &   $Q_L$  & $u_R $ & $d_R $ & $L$  &  $e_R$ & $N_R$ \\
\hline  $q_{Y}$  &  1/6    & - 2/3  & 1/3   &  -1/2   & 1 &  0  \\
\hline   $q_{B}$  & -1    & 0  & 0   & -1   & 0  & 0 \\
\hline \end{tabular}
\end{center}
 \caption{\small Fermion spectrum charges in the $Y$-basis for the Madrid model \cite{Ibanez}.}
\label{charges}
\end{table}

\subsection{Anomalous and anomaly-free regions: numerical results}

\begin{figure}[t]
{\centering \resizebox*{12cm}{!}{\rotatebox{-90}
{\includegraphics{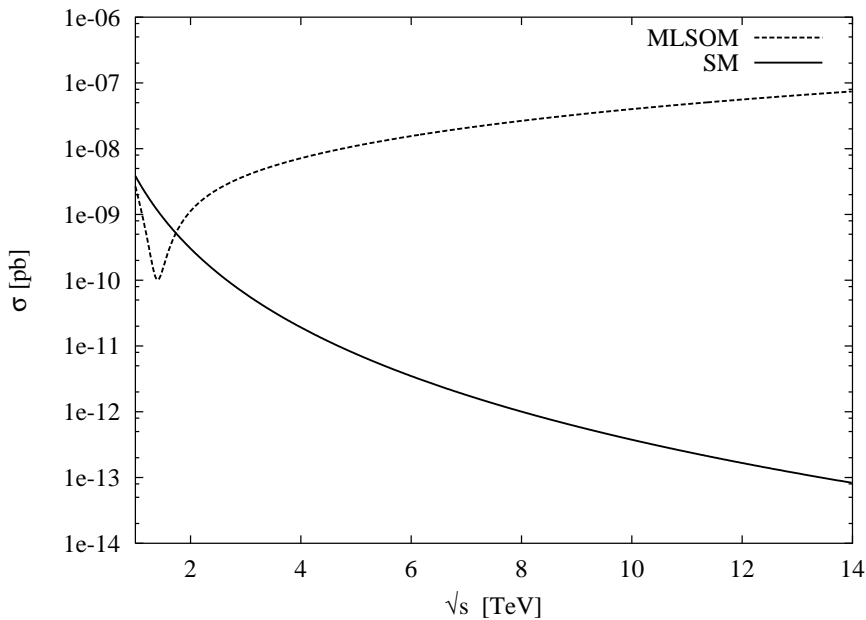}}}\par}
\caption{\small Partonic cross section for the anomalous 
process $gg\rightarrow \g\g$ with $\tan\beta=40$, $g_B=0.1$, $M_{\chi}=10$ GeV and $M_{1}=1$ TeV.
The lines refer to the cross section evaluated for different values of the
center of mass energy $\sqrt{s}$.}
\label{MLSOm_s}
\end{figure}

\begin{figure}[t]
{\centering \resizebox*{12cm}{!}{\rotatebox{-90}
{\includegraphics{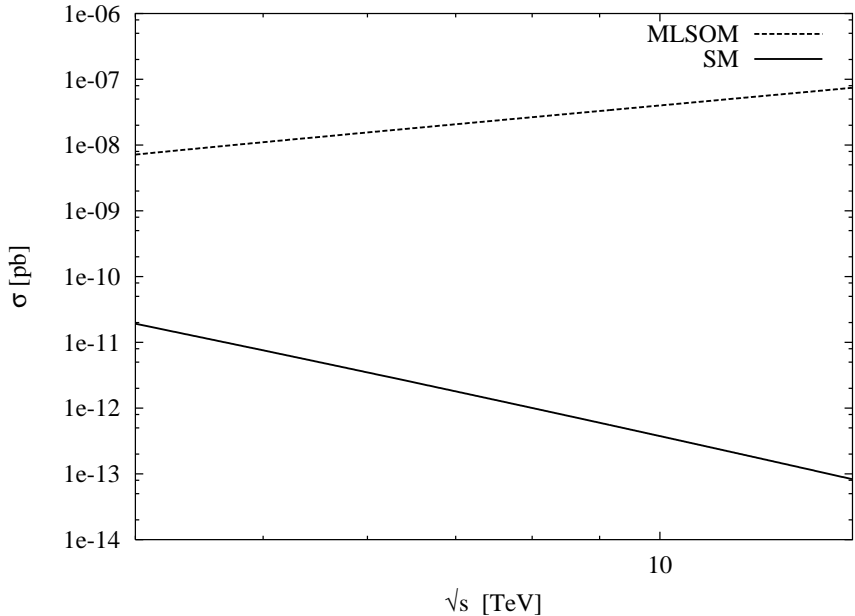}}}\par}
\caption{\small The partonic cross section with a parameter choice as in Fig. \ref{MLSOm_s}.
The dashed upper lines refer to MLSOM cross sections evaluated with
and without the $\chi$ exchange, while the decreasing solid line refers to
the SM.
}
\label{mlsom_sm_14T}
\end{figure}

\begin{figure}[t]
{\centering \resizebox*{12cm}{!}{\rotatebox{-90}
{\includegraphics{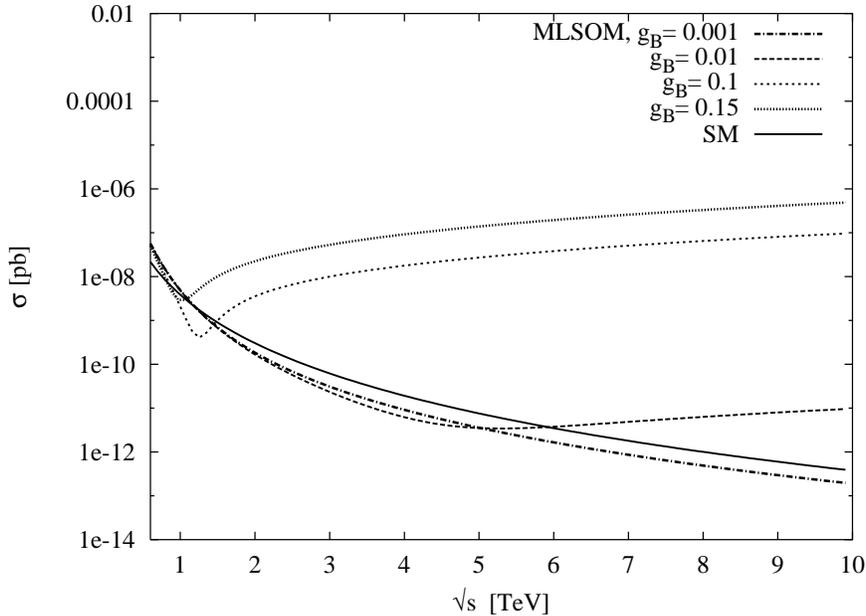}}}\par}
\caption{\small Partonic cross section
plotted for different values of the coupling constant $g_B$. The parameters are chosen as before.
}
\label{mlsom_gB_14T}
\end{figure}

\begin{figure}[t]
{\centering \resizebox*{12cm}{!}{\rotatebox{-90}
{\includegraphics{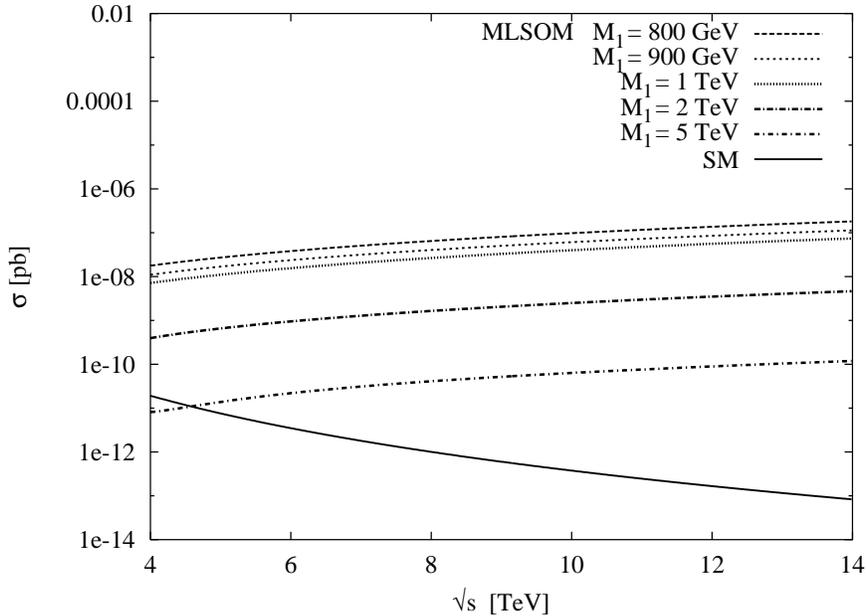}}}\par}
\caption{\small Partonic cross section for the anomalous 
process $gg\rightarrow \g\g$, $\tan\beta=40$, $g_B=0.1$, and $M_{\chi}=10$ GeV
plotted for different values of the St\"uckelberg mass $M_1$.
}
\label{mlsom_M1_14T}
\end{figure}

\begin{figure}[t]
{\centering \resizebox*{12cm}{!}{\rotatebox{-90}
{\includegraphics{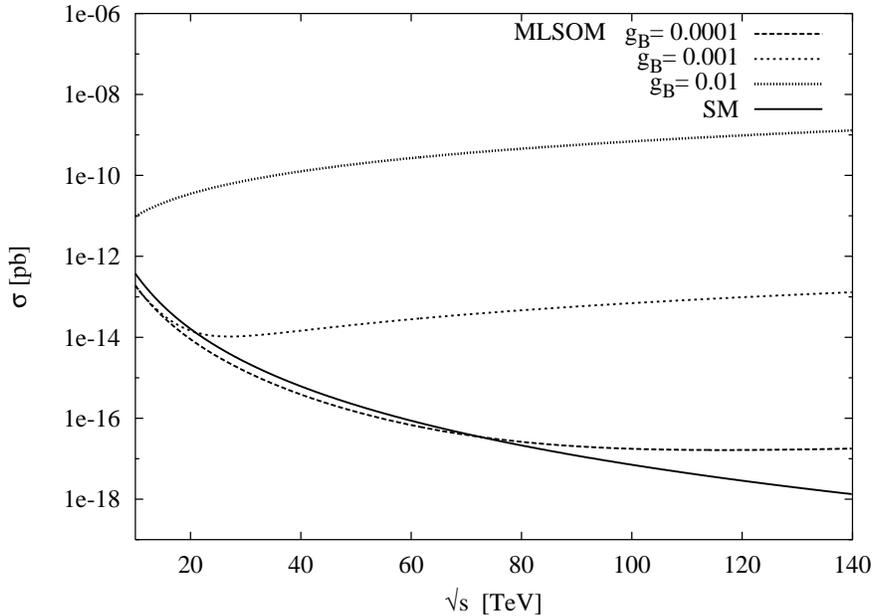}}}\par}
\caption{\small Partonic cross section for the anomalous 
process $gg\rightarrow \g\g$, $\tan\beta=40$ $M_1=800$ 
GeV and $M_{\chi}=10$ GeV. The different plots show a 
comparison between the SM and the MLSOM cross sections,
at very high energies, for small value of the coupling constant $g_B$.
}
\label{mlsom_M1_140T}
\end{figure}

\begin{figure}[t]
{\centering \resizebox*{12cm}{!}{\rotatebox{-90}
{\includegraphics{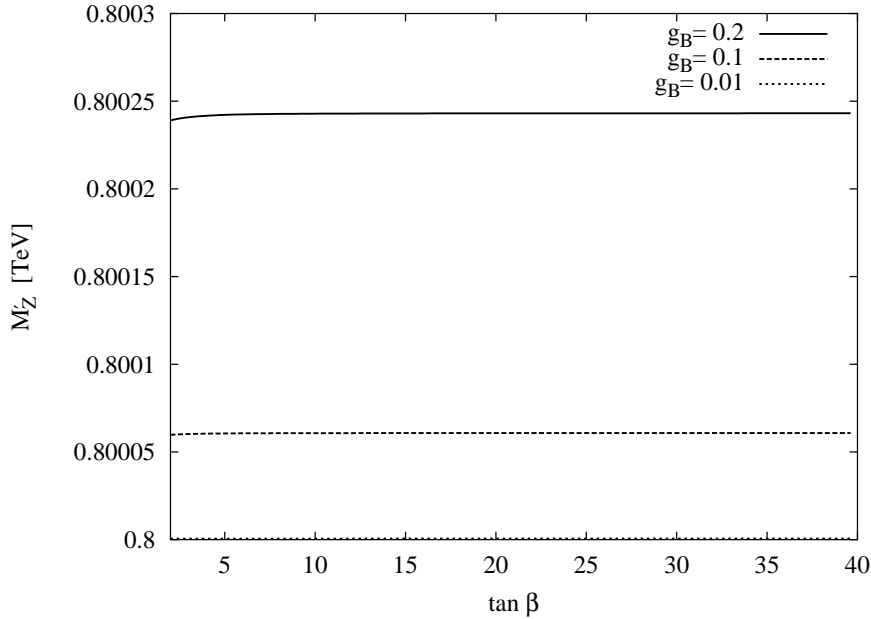}}}\par}
\caption{\small Behavior of the mass of the additional anomalous $Z^{\prime}$
as a function of $\tan{\beta}$ for different values of $g_B$. The variations are very small.}
\label{Mzp_tanbeta}
\end{figure}
\begin{figure}[t]
{\centering \resizebox*{12cm}{!}{\rotatebox{-90}
{\includegraphics{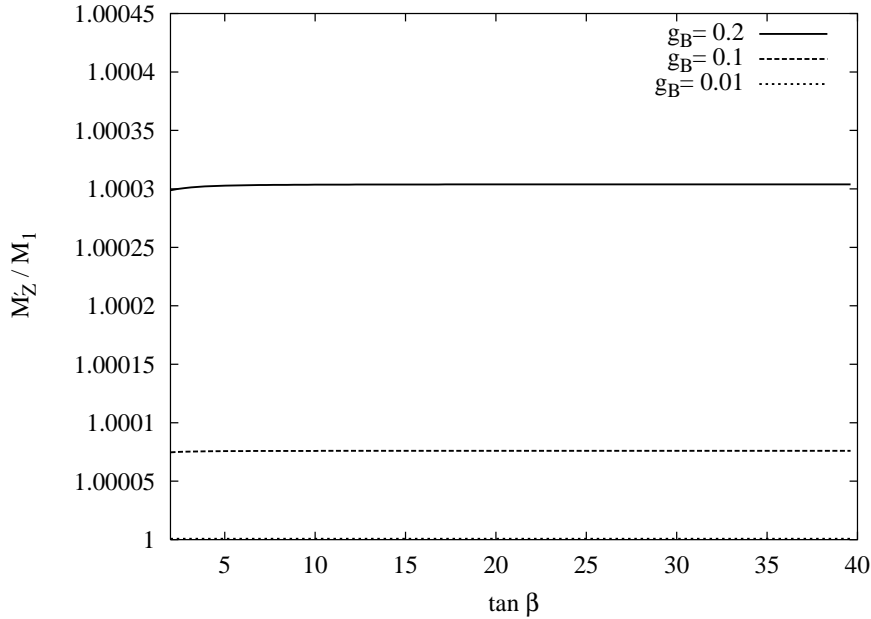}}}\par}
\caption{\small Mass of the extra $Z^\prime$ gauge boson 
for different values of the St\"uckelberg mass.}
\label{M1Mzp}
\end{figure}

\begin{figure}[t]
{\centering \resizebox*{12cm}{!}{\rotatebox{-90}
{\includegraphics{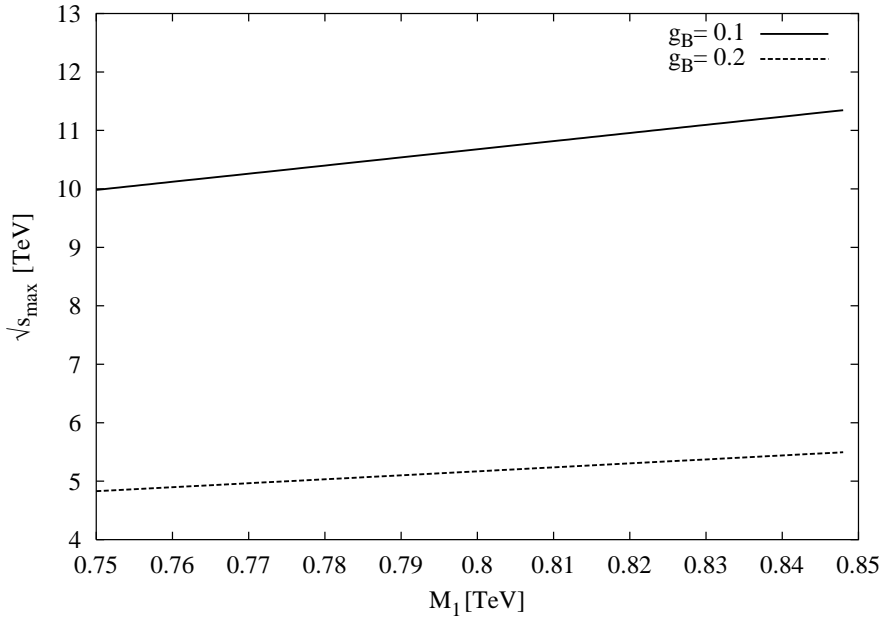}}}\par}
\caption{\small Bound for different values of the St\"uckelberg mass and $\tan{\beta}=40$.
The bound grows as we reduce the anomalous coupling,
and approaches the Standard Model behavior.
For $g_B$ = 0.01 the model exhibits a bound around $\sqrt{s}_{max}$ = 100 TeV (not shown).}
\label{newbound_M1}
\end{figure}

\begin{figure}[t]
{\centering \resizebox*{12cm}{!}{\rotatebox{-90}
{\includegraphics{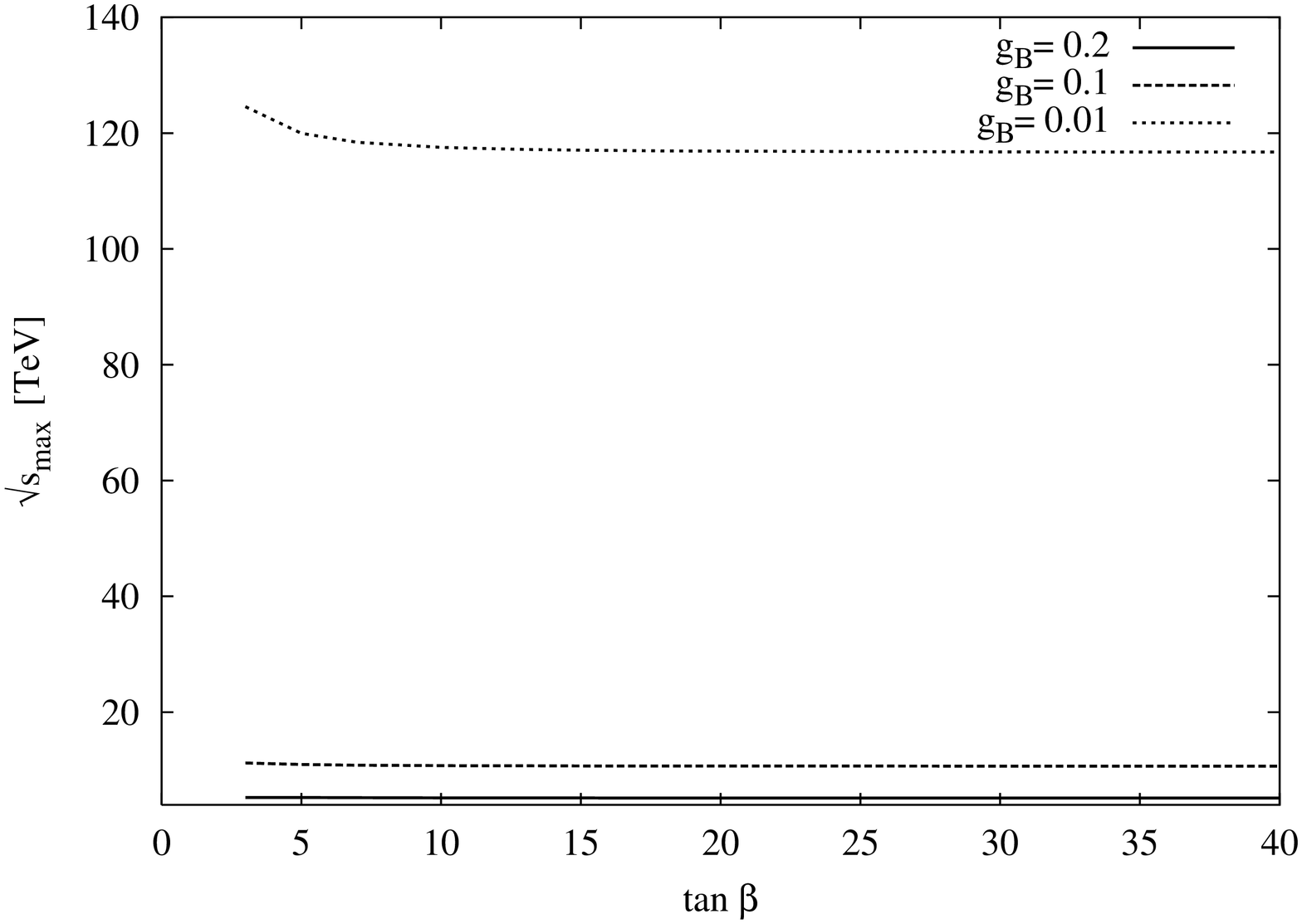}}}\par}
\caption{\small Bound for different values of $\tan\beta$, for $M_1=800$ GeV.}
\label{newbound_tanbeta}
\end{figure}

We have implemented in \textsc{Candia} \cite{CANDIA}
a numerical program that will provide full support for the
experimental collaborations for their analysis of the main 
signals in this class of models at the LHC,
with the charge assignments discussed in the previous section.
\textsc{Candia} has been planned to deal with the analysis of extra
neutral interactions at hadron colliders in specific channels,
such as Drell-Yan processes and double prompt photon processes 
with the highest precision, and it is under intense development. 
The program is entirely based on
the theory developed in \cite{CIK,CIM1,CIM2,CI} and in this work and is tailored
to determine the basic processes, which provide the signal
for the anomalous and anomaly-free extra $Z'$ both from string and GUT models.
The QCD corrections, including the parton evolution, is treated with extreme
accuracy using the theory of the logarithmic expansions, developed in the
last several years. Here we will provide only parton level results for
the anomalous processes that we have discussed in the previous sections,
which clarify the role played by the WZ Lagrangian in the restoration of unitarity at high energy.
A complete analysis for the LHC is under way.

\begin{itemize}
\item{ \bf $\sigma$ reduction by the exchange of $\chi$}

We show in Fig. \ref{MasslessGGgaga} a plot of the (small) 
but increasing partonic cross section
for double prompt photon production from anomalous gluon fusion. 
We have chosen a typical
SM-like value for the coupling constant of the extra $Z'$ included 
in the analysis and varied the center
of mass energy of few GeV's around 4.2 TeV.
We show two plots, both in the brane model, one with the inclusion of
the axi-Higgs $\chi$ and one without it, with only the exchange of the $Z$ and $Z'$.
Notice that the exchange of the $\chi $ is a separate gauge invariant contribution.
We have chosen a St\"uckelberg mass of 800 GeV.
The plots show the theoretically expected reduction of the
linear growth of the cross section, but the improvement is tiny,
for these values of the external parameters.
In these two plots, the fermion masses have been removed,
as we worked in the chiral limit. The inclusion of all the mass
effects in the amplitude has an irrelevant effect on the growth
of the cross section. This is shown in Fig. \ref{MassiveGGgaga} where,
again, the inclusion of the axi-Higgs lowers the growth, but only insignificantly.
We have analyzed the behavior of the cross sections in the
presence of a light (about ten or more GeV's) axi-Higgs, 
but also in this case the effects are negligible.
This feature can be easily checked from Eq. \ref{Mchi}; in fact,
the mass term $M_{\chi}$ is contained in the denominator of the propagator
for the scalar, and in the TeV's region we have $(s-M_{\chi}^2)\approx s$.
The numerical value of the unitarity bound remains essentially unchanged.

\item{\bf  Anomaly-free and anomalous regions}

An interesting behavior shows up in Fig. \ref{SM_vs_MLSOm},
where we compare the results in the SM
and in the MLSOM for the same cross section, starting
at a lower energy. It is clear that, while the SM result, being anomaly-free,
is characterized by a fast-falling cross section,
in the MLSOM it is very different. In particular, one
finds a region of lower energy, where essentially the model
follows the SM behavior (below 1 TeV) - but  smaller by a
factor of 10 - the growth of the anomalous contributions
still being not large; and a region of higher energy, where the
anomalous contributions take over (at about 2 TeV)
and which drive the growth of the cross section, as in the previous
two plots. There is a minimum at about 1.2 TeV, which is the point
at which the anomalous subcomponent becomes sizeable.

\item{\bf $\tan \beta$,  $g_b$ and $M_1$ variations}

The variation of the same behavior shown in the previous
plot with $\tan \beta$ is shown in Fig.\ref{tanbeta_vs_MLSOm},
where we have varied this parameter from small to larger
values (10-50). The depth of the minimum increases as we increase
this value. At the same time, the cross section tends to
fall much steeper, starting from larger values in the anomaly-free region.

In Fig.\ref{gB_vs_MLSOm} we study the variation of $\sigma$
as we tune the coupling of the anomalous $U(1)$, $g_B$.
A very small value of the coupling tends to erase the
anomalous behavior, rendering the anomalous components subleading.
The cross section then falls quite fast before increasing reaching the bound.
The falling region is quite visible for the two values of $g_B=0.001$ and $0.01$,
showing that the set of minimum points, or the anomaly-free region,
is pushed up to several TeV's,
in this case above 4.5 TeV. The unitarity bound is weaker,
being pushed up significantly. The situation is opposite for stronger values of $g_B$.

A similar study is performed in Fig.\ref{diciotto}, but
for a varying St\"uckelberg mass. As this mass parameters
increase, the anomaly-free region tends to grow wider and
the cross section stabilizes. For instance, for a value of
the St\"uckelberg mass around 5 TeV, the region in which $\sigma$
has a normal behavior moves up to 3.5 TeV. The explanation of this
result has to be found in the fact that the anomalous growth is controlled by the
mass of the anomalous $Z'$ in the s-channel, appearing in
the denominator of the cross section. This suppression is seen both
in the direct diagram and in the counterterm diagram, which describes the exchange
of the axi-Higgs. Obviously, it is expected that as we reduce the
coupling of the anomalous gauge boson, the anomalous behavior is reduced as
well.  The different behavior of the cross section in the SM and
MLSOM cases can easily be inferred from
Fig.\ref{MLSOm_s} having chosen a St\"uckelberg
mass of the order of 1 TeV.
A similar behavior is quite evident also from Fig.\ref{mlsom_sm_14T},
from which it appears that in the MLSOM the deviations compared to the SM
partonic predictions get sizeable at parton level already at an energy of 4-6 TeV.
Notice also from Fig.\ref{MassiveGGgaga} that the presence of the axi-Higgs seems
to be irrelevant for the chosen values of the couplings and parameters of the model.
We show in Fig.~\ref{mlsom_gB_14T} a plot of the dependence of the predictions on $g_B$ at
larger energy values, which appears to be quite significant.

Furthermore, in the TeV's region the MLSOM predictions for small values of
the coupling constant ($g_B=0.001$) go below the SM prediction and
this is due to the axi-higgs exchange, which is negative in this kinematical domain.
Moving below to 1 TeV the axi-higgs interference has an opposite sign and
the MLSOM predictions are above the SM.
A similar analysis, this time for a varying St\"uckelberg mass,
is shown in Fig.\ref{mlsom_M1_14T}, and also in this case, as in the previous one,
the results confirm that this dependence is very relevant.

Finally, in Fig.\ref{mlsom_M1_140T} we plot the SM and MLSOM 
results on a larger interval, from 10 to 140 TeV,
from which the drastically different behavior of the two cross 
sections are quite clear. Notice that as we lower $g_B$, for 
instance down to $10^{-4}$, the anomaly-free region extends up 
to energy values that are of the order of 200 TeV or so. 
We conclude that the enhancement of the anomalous contributions with respect to the SM
prediction are in general quite large and very sensitive to the mass
of the extra $Z'$ and to the strength of the anomalous coupling.
Interestingly, a very weakly coupled $Z'$ gives a cross section that has
a faster fall-off compared to the SM case in the anomaly-free region. 
The MLSOM and SM predictions
intersect at a very large energy scale (140 TeV), when the 
anomalous contribution starts to increase.

Before drawing conclusions, it is necessary to comment on the other dependence,
that on $\tan\beta$, which appears to be far less significant compared 
to $M_1$ (or $M_Z'$) and $g_B$. This third parameter essentially has a
(very limited) influence on the mass of the extra $Z'$ and on the overall
predictions. This is clearly shown in
Figs.\ref{Mzp_tanbeta} and \ref{M1Mzp}, where we have varied both 
$\tan\beta$ and $g_B$. Therefore, the mass of the extra $Z'$ and the
St\"uckelberg mass may be taken to be essentially coincident, to a first approximation.

\item{\bf The bounds}

We conclude our analysis with two plots, depicted 
in Figs.\ref{newbound_M1} and \ref{newbound_tanbeta}, 
which show the variations of the bounds with the parameters
$g_B$ and $M_1$. In the first plots, shown in Fig.\ref{newbound_M1}, 
we choose a large value of $\tan\beta$ and we have varied both the 
St\"uckelberg mass and the strength of the anomalous coupling. 
For a St\"uckelberg mass around 1 TeV, the bound is around 5 TeV, 
for an anomalous coupling $g_B=0.2$. For a smaller value of $g_B=0.1$ 
the bound grows to 10 TeV. For a smaller value of $g_B=10^{-2}$, 
the bound is around 100-120 TeV. This result is particularly interesting, 
because it should allow one to set limits on the St\"uckelberg mass and the
value of the anomalous couplings at the LHC in the near future. 
Smaller values of $g_B$ are tested in Fig.\ref{newbound_tanbeta}, 
where the bound is shown to increase significantly as $g_B$ gets smaller.

\end{itemize}
\section{Conclusions}

We have analyzed the connection between the WZ term and the GS 
mechanism in the context first of simple models and then 
in a complete brane model, containing three extra anomalous
$U(1)$. We have shown that the WZ method of cancellation 
of the anomaly does not protect
the theory from an excessive growth, which is bound to
violate unitarity beyond a certain
scale. We have also studied the connection between the two 
mechanisms, illustrating the corresponding differences.

We have quantified the unitarity bound for several choices of the parameters
of the theory. The significant dependences are those on the St\"uckelberg
mass and the coupling constants of the anomalous generators. 
We have also shown that the exchange of a physical axion lowers
the cross section, but not significantly, whose growth remains
essentially untamed at high energy. We have shown that in these
models one can identify a region that is SM-like, where some 
anomalous processes have a fast fall-off, from a
second region, where the anomaly dominates. Clearly, more investigations
are necessary in order to come out with more definitive predictions 
for the detection of anomalous interactions at the LHC, since our 
analysis has been confined to the parton level. The approach that we
have suggested, the use of BIM amplitudes to search for unitarity 
violations at future colliders, can be a way in the
search to differentiate between non-anomalous 
\cite{Langacker} and anomalous extra $Z'$.
Our objective here has been to
show that there is a systematic way to analyze the two mechanism
for canceling the
anomalies at the phenomenological level and that unitarity issues
are important in order to characterize the region in which a
certain theory starts to be dominated by the chiral anomaly.
We hope to address these points in the near future in related work.

\centerline{\bf Acknowledgements}

We thank Alan R. White, Nikos Irges, Theodore Tomaras and Marco Roncadelli for discussions.
C.C. thanks Theodore Tomaras and Elias Kiritsis
for hospitality at the Univ. of Crete.
The work of C.C. was supported (in part) by the European Union through the Marie Curie Research
and Training Network ``Universenet'' (MRTN-CT-2006-035863) and by The Interreg II Crete-Cyprus Program.

\section{Appendix. The cross section }
The total cross section is given by the sum of all the contributions shown in Fig.\ref{GS}
\ba
\sigma^{MLSOM}(s)&=&\sigma_A(s) + \sigma_{B}(s) + \sigma_{C}(s)+  \sigma_{D}(s) +  \sigma_{E}(s) +  \sigma_{F}(s)  \nonumber\\
&& + \, \sigma_{AB}(s) + \sigma_{AE}(s) +  \sigma_{AF}(s)+  \sigma_{BF}(s) +  \sigma_{EF}(s).
\label{totalcross}
\ea

The interference term between $Z$ exchange and $\chi$ exchange, diagram (b) in Fig. (\ref{GS}), is given by
\ba
\sigma_{AB}(s)= \frac{1}{16 \pi}
\left[\sum_q \frac{1}{2} c_1^{q}A_{6,q}\right]\left[\sum_{f} \frac{1}{2} c_2^{f}A_{6,f}\right]
g^{\chi}_{gg}g^{\chi}_{\g\g}\frac{s^4}{M_Z^2(s-M_{\chi}^2)}\,.
\ea
The following interference terms are vanishing 
\ba
&&{\cal M}_A {\cal M}_C^\dagger + {\cal M}_C {\cal M}_A^\dagger = 0,  \qquad 
{\cal M}_A {\cal M}_D^\dagger + {\cal M}_D {\cal M}_A^\dagger = 0,   \nonumber\\
&&{\cal M}_B {\cal M}_C^{\dagger}  + {\cal M}_C {\cal M}_B^\dagger = 0,   \qquad
{\cal M}_B {\cal M}_D^{\dagger} + {\cal M}_D {\cal M}_B^{\dagger} = 0.
\ea
The interference term between the exchange of the $Z$ boson and $\chi$ exchange, 
diagram (e) in Fig. (\ref{GS}), gives
\ba
\sigma_{AE}(s)& =&   \frac{s^{4}}{1024 M^{2}_{Z} \pi (s - M^{2}_{\chi}) }
\left[ \sum_q \frac{1}{2} c_{1}^{q}  A_{6,q}  \right]
\left[  \sum_{f} \frac{1}{2} c_{2}^{f}   A_{6,f}  \right]   \nonumber\\
&& \times \left[\sum_{ f^\prime } C_{0}(s, m_{ f^\prime}) c^{\chi,  f^\prime}_{\g\g}   \right]
\left[ \sum_{ q^\prime} C_{0}(s, m_{ q^\prime}) c^{\chi, q^\prime}_{gg}   \right],
\ea
while the interference term between $\chi$ exchange and $Z^\prime$ boson exchange contributes with
\ba
\sigma_{BE}(s) =- \frac{ s^3}{16 \pi (s - M_{\chi}^{2})^{2}} g^{\chi}_{\gamma \gamma}  g^{\chi}_{gg}
\left[ \sum_{ f} C_{0}(s, m_{ f}) c^{\chi, f}_{\g\g} \right]
\left[\sum_ q C_{0}(s, m_q) c^{\chi, q}_{gg}  \right].
\ea
The other interference term in the cross section is given by
\ba
\sigma_{CD}(s) = \frac{ s^3}{16 \pi (s - M_{\chi}^{2})^{2}}
g^{\chi}_{\gamma \gamma} g^{\chi}_{gg}     \left[ \sum_{q} C_{0}(s, m_{q} ) c^{\chi, q}_{gg}  \right]
 \left[ \sum_f C_{0}(s, m_f) c^{\chi, f}_{\g\g}    \right],
\ea
 so we obtain
\ba
\sigma_{CD}(s) + \sigma_{BE}(s) = 0.
\ea
Other interference terms also vanish; in fact, we get
\ba
{\cal M}_C {\cal M}_E^{\dagger} + {\cal M}_E {\cal M}_C^\dagger = 0, \qquad
{\cal M}_D {\cal M}_E^{\dagger} + {\cal M}_E {\cal M}_D^{\dagger} = 0,
\ea
\ba
{\cal M}_C {\cal M}_F^{\dagger} + {\cal M}_F {\cal M}_C^\dagger = 0, \qquad
{\cal M}_D {\cal M}_F^{\dagger} + {\cal M}_F {\cal M}_D^{\dagger} = 0.
\ea
The interference term between $Z$ exchange and $Z^\prime$ exchange takes the form
\ba
\sigma_{AF} = \frac{1}{1024 \pi} \left[\sum_q \frac{1}{2} {d}_1^{\,q}A_{6,q}\right]
\left[\sum_{f} \frac{1}{2} d_2^{f}A_{6,f}\right]  \left[\sum_{q^\prime} \frac{1}{2} c_1^{q^\prime}A_{6,q^\prime}\right]
\left[\sum_{f^\prime} \frac{1}{2} c_2^{f^\prime}A_{6,f^\prime}\right]
\frac{s^5}{M_{Z}^{2}\, M_{Z^\prime}^2}\,.
\ea
The interference term between $Z^\prime$ exchange and $\chi$ exchange 
of diagram (b) in Fig. (\ref{GS}) is given by
\ba
\sigma_{BF}(s)=\frac{1}{16\pi}
\left[\sum_q \frac{1}{2} d_1^{\,q}A_{6,q}\right]\left[\sum_{f} \frac{1}{2} d_2^{f}A_{6,f}\right]
g^{\chi}_{gg}g^{\chi}_{\g\g}\frac{s^4}{M_{Z^\prime}^2(s-M_{\chi}^2)}\,.
\ea
Finaly, the interference between $Z^\prime$ exchange and $\chi$ exchange, 
diagram (e) in Fig. (\ref{GS}), gives
\ba
\sigma_{EF}(s)& =&   \frac{s^{4}}{1024\, M^{2}_{Z^\prime}\, \pi \, (s - M^{2}_{\chi}) }
\left[ \sum_q \frac{1}{2} d_{1}^{\,q}  A_{6,q}  \right]
\left[  \sum_{f} \frac{1}{2} d_{2}^{f}   A_{6,f}  \right]   \nonumber\\
&&  \times \left[\sum_{ f^\prime } C_{0}(s, m_{ f^\prime}) c^{\chi,  f^\prime}_{\g\g}   \right]
\left[ \sum_{ q^\prime} C_{0}(s, m_{ q^\prime}) c^{\chi, q^\prime}_{gg}   \right].
\ea

\end{document}